\documentclass[twocolumn,english,prl,aps,superscriptaddress,notitlepage,twocolumn]{revtex4-1}
\usepackage[T1]{fontenc}
\usepackage[latin9]{inputenc}
\usepackage{array}
\usepackage{booktabs}
\usepackage{amsmath}
\usepackage{graphicx}

\makeatletter

\providecommand{\tabularnewline}{\\}

\usepackage[colorlinks,citecolor=blue,linkcolor=blue]{hyperref}
\usepackage{booktabs}

\usepackage{array}
\usepackage[table,xcdraw]{xcolor}
\usepackage{multirow}

\makeatother

\usepackage{babel}
\begin{document}
\title{Hierarchical Coarse-grained Approach to the Duration-dependent Spreading
Dynamics on Complex Networks}
\date{\today}
\author{Jin-Fu Chen}
\address{Beijing Computational Science Research Center, Beijing 100193, China}
\address{Graduate School of China Academy of Engineering Physics, Beijing,
100193, China}
\author{Yi-Mu Du}
\address{Graduate School of China Academy of Engineering Physics, Beijing,
100193, China}
\author{Hui Dong}
\email{hdong@gscaep.ac.cn}

\address{Graduate School of China Academy of Engineering Physics, Beijing,
100193, China}
\author{Chang-Pu Sun}
\email{cpsun@csrc.ac.cn}

\address{Beijing Computational Science Research Center, Beijing 100193, China}
\address{Graduate School of China Academy of Engineering Physics, Beijing,
100193, China}
\begin{abstract}
Various coarse-grained models have been proposed to study the spreading
dynamics on complex networks. A microscopic theory is needed to connect
the spreading dynamics with individual behaviors. In this letter,
we unify the description of different spreading dynamics by decomposing
the microscopic dynamics into two basic processes, the aging process
and the contact process. A hierarchical duration coarse-grained (DCG)
approach is proposed to study the duration-dependent processes. Applied
to the epidemic spreading, such formalism is feasible to reproduce
different epidemic models, e.g., the SIS and the SIR models, and to
associate the macroscopic spreading parameters with the microscopic
mechanism. The DCG approach enables us to study the steady state of
the duration-dependent SIS model. The current hierarchical formalism
can also be used to describe the spreading of information and public
opinions, or to model a reliability theory on networks.
\end{abstract}
\maketitle
\narrowtext

\textit{Introduction.---}The epidemics \citep{Moore2000,Pastor-Satorras2001,PastorSatorras2001,May2001,Cai2016,Hindes2016},
rumors or information \citep{GOFFMAN1964,DALEY1964,Moreno2004,Nematzadeh2014,Gleeson2016},
and public opinions \citep{SZNAJD-WERON2000,Dornic2001,Krapivsky2003,Fernandez-Gracia2014},
etc., usually spread on complex networks with predefined structures.
The spreading dynamics is strongly affected by the characteristic
of the structural networks \citep{Albert2002,MarcBarthelemy2012}.
The utilization of the susceptible-infected-susceptible (SIS) and
the susceptible-infected-recovered (SIR) models initiated the study
of the epidemic spreading on networks \citep{Pastor-Satorras2001,May2001}.
The network structure, known as the degree distribution, affects the
epidemic threshold \citep{Chakrabarti2008,Mieghem2009,Gomez2010,Castellano2010,Ferreira2012,Li2012,MarcBarthelemy2012,Goltsev2012,Lee2013,Boguna2013,Castellano2017,Parshani2010,Wei2020},
which is an index to determine the epidemic phase transition whether
the disease spreads over society. The spreading dynamics is also affected
by the microscopic mechanism, namely, the rules of the state change
and the transition rates of the basic processes. Currently, a unified
spreading model combining both the network structure and microscopic
mechanism remains missing. In this Letter, we propose a unified formalism
to describe the spreading dynamics on the network with general microscopic
mechanism.

For the Markovian spreading models with constant transition rates,
serial mean-field theories have been proposed to describe the spreading
dynamics with neglecting the correlation between nodes \citep{Dorogovtsev2008,Castellano2009,PastorSatorras2015}.
For instance, the epidemic threshold of the standard SIS on networks
was obtained via the heterogeneous-mean-field approach with the degree
distribution \citep{Pastor-Satorras2001,PastorSatorras2001,May2001,MarcBarthelemy2012},
and was later refined via the quenched-mean-field approach by considering
the details of the network topology \citep{Chakrabarti2008,Mieghem2009,Castellano2010,Gomez2010,Ferreira2012}.
For a real-world epidemic, the transmissibility varies in different
disease stages \citep{Hoppensteadt1974,Feng2007,Magal2010,Liu2015,Wang2016,Mieghem2013,Cator2013,Yang2016,Chen2018,Mieghem2019,Arruda2020,Starnini2017}.
Namely, the infection rate relies on the infection duration. Such
non-Markovian property was proposed to dramatically affect the spreading
dynamics and alter the epidemic threshold \citep{Mieghem2013,Cator2013,Yang2016,Chen2018,Mieghem2019,Arruda2020}.
Here we extend the mean-field theories to the duration-dependent spreading
models by introducing the probability density function (PDF) of the
duration with the varied transition rates adopted from the reliability
theory \citep{SystemReliabilityTheory,2006a,Rocchi2017}. In our formalism,
the spreading dynamics are decomposed into two basic processes, the
aging process describing the self-evolution of one node (single-body
process), and the contact process describing the state change of two
connected nodes (two-body process). The two processes are modeled
here as a continuous-time stochastic process among a set of discrete
states.

Inspired by the coarse-grained approaches of the complex networks
\citep{Gfeller2007,Gfeller2008,Chen2010,Shen2011}, the duration-dependent
spreading models are presented in three hierarchies, the microscopic,
the mesoscopic, and the macroscopic models. In the microscopic model,
we derive the basic equations of the PDF of each node with neglecting
the correlation between nodes. In the microscopic model, a duration
coarse-grained (DCG) approach is proposed to obtain the coarse-grained
PDF of the ensemble with the same degree, and gives a refined spreading
rate for the duration-dependent SIS model. The microscopic and the
mesoscopic models extend the quenched and the heterogeneous mean-field
approaches to the duration-dependent spreading models, respectively.
The macroscopic model describes the spreading dynamics by assuming
the identical PDF of all nodes, and recovers to the compartmental
epidemic model \citep{1927a,Brauer2019,Du_2020reliability}. The macroscopic
model is quantitatively applicable for a homogeneous network with
a narrow degree distribution, but gives qualitative prediction about
the spreading dynamics.

\begin{figure}
\includegraphics[width=5.5cm]{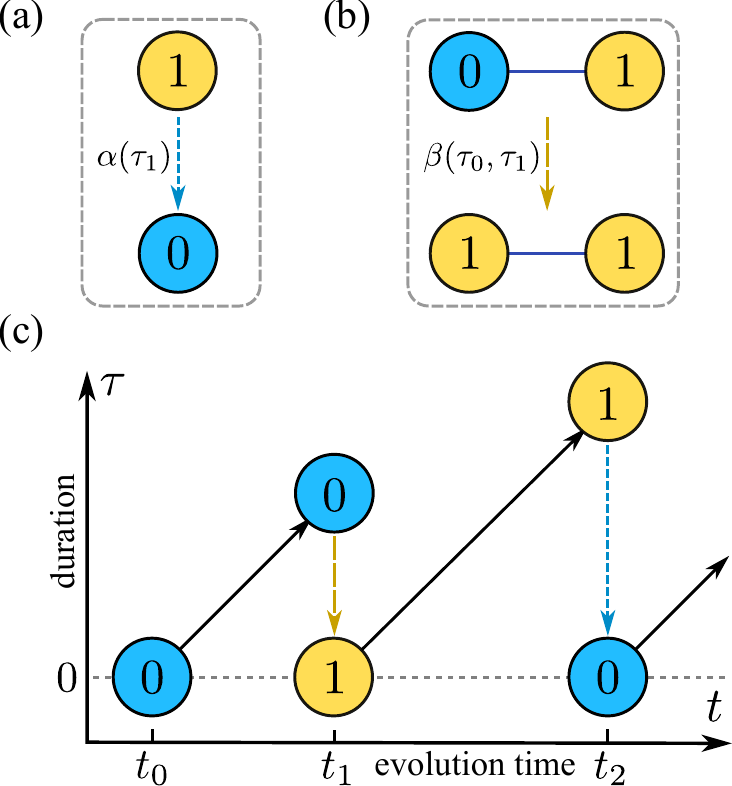}

\caption{(Color online) Diagrams of the two basic processes. (a) The aging
process $1\protect\overset{\alpha(\tau_{1})}{\protect\longrightarrow}0$
describes the recovery of an infected node. (b) The contact process
$0+1\protect\overset{\beta(\tau_{0},\tau_{1})}{\protect\longrightarrow}1+1$
describes the infection of a susceptible node raised by an linked
node. (c) The evolution of one node, where $\tau$ and $t$ are the
duration on one state and the evolution time, respectively. \label{fig:Diagrams-of-two}}
\end{figure}

\textit{Two basic processes.---}We consider an undirected network
with $N_{T}$ nodes represented by an adjacency matrix $A_{lm}$.
The node state is picked from the state set $i\in\{0,1,2,...\}$.
The state evolution is governed by two basic processes, the aging
process and the contact process, as shown in Fig. \ref{fig:Diagrams-of-two}(a)
and (b), respectively.

The aging process describes the state change $i\overset{\alpha_{i^{\prime},i}}{\longrightarrow}i^{\prime}$
of one single node. The transition rate $\alpha_{i^{\prime},i}(\tau_{i})$
generally relates to its duration $\tau_{i}$ on the state $i$ \citep{2006a}.
The maximum entropy principle can be used to estimate the most probable
transition rate \citep{Du2020,Du_2020reliability}, when limited information,
e.g., the mean infection time, is known about the process.

The contact process describes the correlated state change $i+j\overset{\beta_{i^{\prime}j^{\prime},ij}}{\longrightarrow}i^{\prime}+j^{\prime}$
of two linked nodes. The transition rate $\beta_{i^{\prime}j^{\prime},ij}(\tau_{i},\tau_{j})$
relates to the duration $\tau_{i}$ and $\tau_{j}$ of the two nodes
in the states $i$ and $j$. Different patterns exist for the contact
process, e.g., the exchange process $i+j\overset{\beta_{ji,ij}}{\longrightarrow}j+i$
and the infection process $i+j\overset{\beta_{jj,ij}}{\longrightarrow}j+j$.

A majority of spreading models can be constructed with the two basic
processes above. For example, two states $0$ and $1$ are the susceptible
and the infected states in the SIS model. The basic processes are
an aging process $1\overset{\alpha(\tau_{1})}{\longrightarrow}0$
with the recovery rate $\alpha(\tau_{1})$, and a contact process
$0+1\overset{\beta(\tau_{0},\tau_{1})}{\longrightarrow}1+1$ with
the infection rate $\beta(\tau_{0},\tau_{1})$. The duration-dependent
infection rate $\beta(\tau_{0},\tau_{1})$ reflects the change of
both the vulnerability of the susceptible state and the transmissibility
of the infected state with their duration. The typical evolution of
one node is shown in Fig. \ref{fig:Diagrams-of-two}(c). At the initial
time $t=t_{0}$, the node stays in the state $0$ with zero duration
$\tau=0$. Its state changes accompanied with resetting the duration
at time $t_{1}$ and $t_{2}$ due to the contact and the aging processes.
In the typical model of rumor spreading \citep{DALEY1964,Moreno2004},
three states $0$, $1$ and $2$ are the ignorant, spreading, and
stifling states, the change of which is governed by three basic processes
$0+1\overset{\beta_{1}(\tau_{0},\tau_{1})}{\longrightarrow}1+1$,
$1+1\overset{\beta_{2}(\tau_{1},\tau_{1})}{\longrightarrow}2+1$,
and $1+2\overset{\beta_{3}(\tau_{1},\tau_{2})}{\longrightarrow}2+2$.
The transition rates generally depend on the duration, but such duration-dependent
effects have seldom been considered in the current studies.

\begin{figure}
\includegraphics[width=6cm]{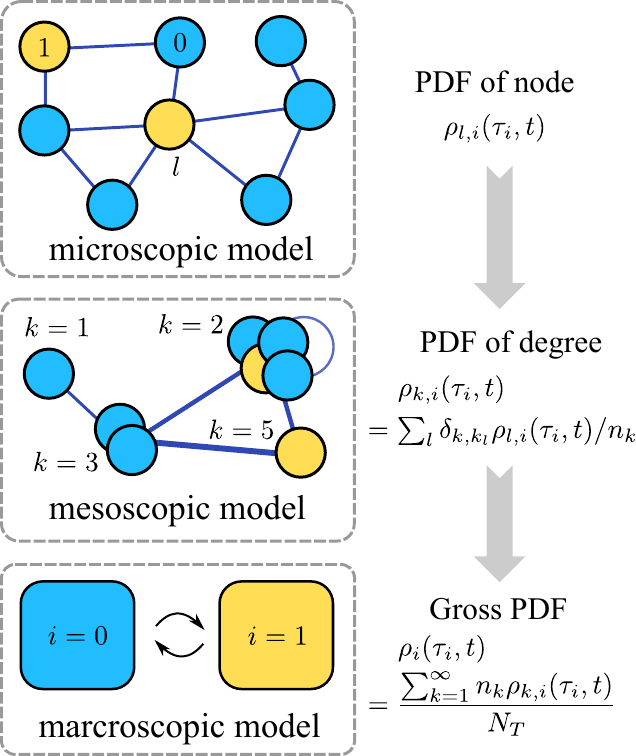}

\caption{(Color online) Hierarchy of the microscopic, the mesoscopic, and the
macroscopic models of the spreading dynamics. The information of the
duration distribution is recorded by the probability density function
$\rho_{l,i}(\tau_{i},t)$, $\rho_{k,i}(\tau_{i},t)$ and $\rho_{i}(\tau_{i},t)$
at different coarse-grained levels. \label{fig:generalspreading-model-at}}
\end{figure}

\textit{Duration-dependent spreading models.---}The conventional
spreading models \citep{Pastor-Satorras2001,MarcBarthelemy2012} with
only recording the node states are not enough to describe the spreading
dynamics with the duration-dependent transition rates. In Fig. \ref{fig:generalspreading-model-at}(a),
we introduce the probability density function (PDF) $\rho_{l,i}(\tau_{i},t)$
of the duration for the node $l$ in the microscopic model. The probability
of the node $l$ in the state $i$ follows as $P_{l,i}(t)=\int_{0}^{\infty}\rho_{l,i}(\tau_{i},t)d\tau_{i}$.
By neglecting the correlation between nodes, the state of the network
is described by the PDF $\rho_{l,i}(\tau_{i},t)$. The equation of
the PDF reads (see the derivation in supplementary materials \citep{supplementarymaterials})

\begin{equation}
\frac{\partial\rho_{l,i}(\tau_{i},t)}{\partial\tau_{i}}+\frac{\partial\rho_{l,i}(\tau_{i},t)}{\partial t}=-\Gamma_{l,i}(\tau_{i},t)\rho_{l,i}(\tau_{i},t).\label{eq:differentialequationofPDF}
\end{equation}
The total transformation rate for the node $l$ of leaving the state
$i$ is $\Gamma_{l,i}(\tau_{i},t)=\sum_{i^{\prime}}\gamma_{l,i^{\prime}i}(\tau_{i},t)$,
with the transformation rate $\gamma_{l,i^{\prime}i}(\tau_{i},t)$
from the state $i$ to the state $i^{\prime}$ explicitly as

\begin{equation}
\gamma_{l,i^{\prime}i}=\alpha_{i^{\prime},i}+\sum_{m,j,j^{\prime}}A_{lm}\int_{0}^{\infty}\beta_{i^{\prime}j^{\prime},ij}\rho_{m,j}d\tau_{j}.\label{eq:gammaliprimei}
\end{equation}
The connecting condition for the PDF at the boundary $\tau_{i}=0$
is determined by the flux to the state $i$ as $\rho_{l,i}(0,t)=\Phi_{l,i}(t)=\sum_{i^{\prime}}\phi_{l,ii^{\prime}}(t),$
where $\phi_{l,ii^{\prime}}(t)=\int_{0}^{\infty}\gamma_{l,ii^{\prime}}(\tau_{i^{\prime}},t)\rho_{l,i^{\prime}}(\tau_{i^{\prime}},t)d\tau_{i^{\prime}}$
is the probability of the node $l$ transforming from the state $i^{\prime}$
to the state $i$ in unit time.

To effectively describe the spreading dynamics without considering
the state of each node, we propose a duration coarse-grained (DCG)
approach to study the duration-dependent effect with the coarse-grained
PDF. In the mesoscopic model, the nodes are sorted into different
ensembles with the degree $k$, as shown in Fig \ref{fig:generalspreading-model-at}(b).
The states of the network are described by the coarse-grained PDF
of the $k$-degree nodes as $\rho_{k,i}(\tau_{i},t)=\sum_{l}\delta_{k,k_{l}}\rho_{l,i}(\tau_{i},t)/n_{k}$
with the population $n_{k}$ of all $k$-degree nodes. The population
of the $k$-degree nodes in the state $i$ follows as $n_{k,i}(t)=n_{k}\int_{0}^{\infty}\rho_{k,i}(\tau_{i},t)d\tau_{i}$.
The PDFs of the nodes with the same degree are assumed identical $\rho_{l,i}(\tau_{i},t)=\rho_{k_{l},i}(\tau_{i},t)$,
and the transformation rate of a node only relies on its degree as
$\gamma_{k_{l},i^{\prime}i}(\tau_{i},t)$. The equation of the coarse-grained
PDF of the $k$-degree nodes is obtained from Eq. (\ref{eq:differentialequationofPDF})
as

\begin{equation}
\frac{\partial\rho_{k,i}(\tau_{i},t)}{\partial\tau_{i}}+\frac{\partial\rho_{k,i}(\tau_{i},t)}{\partial t}=-\Gamma_{k,i}(\tau_{i},t)\rho_{k,i}(\tau_{i},t).\label{eq:differentialequationofDDF1}
\end{equation}
The total transformation rate is $\Gamma_{k,i}(\tau_{i},t)=\sum_{i^{\prime}}\gamma_{k,i^{\prime}i}(\tau_{i},t)$,
and the transformation rate $\gamma_{k,i^{\prime}i}(\tau_{i},t)$
is simplified as

\begin{equation}
\gamma_{k,i^{\prime}i}=\alpha_{i^{\prime},i}+k\sum_{j,j^{\prime}}\sum_{k^{\prime}=1}^{\infty}P(k^{\prime}|k)\int_{0}^{\infty}\beta_{i^{\prime}j^{\prime},ij}\rho_{k^{\prime},j}d\tau_{j},\label{eq:gammak,i'i}
\end{equation}
where the degree correlation $P(k^{\prime}|k)$ describes the degree
distribution of a neighbor of a $k$-degree node, and is determined
by the adjacency matrix $A_{lm}$ as $P(k^{\prime}|k)=\sum_{l,m}\delta_{k,k_{l}}\delta_{k^{\prime},k_{m}}A_{lm}/(kn_{k})$
\citep{supplementarymaterials}. The connecting condition for the
coarse-grained PDF is $\rho_{k,i}(0,t)=\Phi_{k,i}(t)=\sum_{i^{\prime}}\phi_{k,ii^{\prime}}(t),$
where $\phi_{k,ii^{\prime}}(t)=\int_{0}^{\infty}\gamma_{k,ii^{\prime}}(\tau_{i^{\prime}},t)\rho_{k,i^{\prime}}(\tau_{i^{\prime}},t)d\tau_{i^{\prime}}$
is the flux of one $k$-degree node transforming from the state $i^{\prime}$
to the state $i$. An example with explicit equations of PDFs in the
duration-dependent SIS model can be found in the supplementary materials
\citep{supplementarymaterials} or in Ref. \citep{Yang2016}.

At the macroscopic level, a further coarse-grained procedure introduces
the gross PDF $\rho_{i}(\tau_{i},t)=[\sum_{k=1}^{\infty}n_{k}\rho_{k,i}(\tau_{i},t)]/N_{T}$
of all nodes to simplify the spreading dynamics, as shown in Fig.
\ref{fig:generalspreading-model-at}(c). The dynamics is then regarded
to be homogeneous for all nodes independent of the degree. The population
of the nodes in the state $i$ follows as $N_{i}(t)=N_{T}\int_{0}^{\infty}\rho_{i}(\tau_{i},t)d\tau_{i}$.
This approximation is suitable for the homogeneous network with similar
degrees for all nodes. The equation of the gross PDF is obtained from
Eq. (\ref{eq:differentialequationofDDF1}) as

\begin{equation}
\frac{\partial\rho_{i}(\tau_{i},t)}{\partial\tau_{i}}+\frac{\partial\rho_{i}(\tau_{i},t)}{\partial t}=-\Gamma_{i}(\tau_{i},t)\rho_{i}(\tau_{i},t).
\end{equation}
The total transformation rate is $\Gamma_{i}(\tau_{i},t)=\sum_{i^{\prime}}\gamma_{i^{\prime}i}(\tau_{i},t)$,
with the transformation rate $\gamma_{i^{\prime}i}(\tau_{i},t)$ explicitly
as

\begin{equation}
\gamma_{i^{\prime}i}=\alpha_{i^{\prime},i}+\left\langle k\right\rangle \sum_{j,j^{\prime}}\int_{0}^{\infty}\beta_{i^{\prime}j^{\prime},ij}\rho_{j}d\tau_{j}.\label{eq:gammak,i'i-1}
\end{equation}
The effect of the network structure on the spreading dynamics is reflected
by the average degree $\left\langle k\right\rangle =\sum_{k=1}^{\infty}kP(k)$.
The connecting condition for the gross PDF is $\rho_{i}(0,t)=\Phi_{i}(t)=\sum_{i^{\prime}}\phi_{ii^{\prime}}(t)$
with the gross flux $\phi_{ii^{\prime}}(t)=\int_{0}^{\infty}\gamma_{ii^{\prime}}(\tau_{i^{\prime}},t)\rho_{i^{\prime}}(\tau_{i^{\prime}},t)d\tau_{i^{\prime}}$.
Details of the coarse-grained procedures are shown in the supplementary
materials \citep{supplementarymaterials}.

Our spreading models can be widely used to describe different problems
with different meanings of the states and the nodes. For example,
the node states describe disease of individuals in an epidemic model
\citep{Pastor-Satorras2001}, or performance of components in a reliability
model \citep{Rocchi2017}. The transformation rates and the connecting
conditions are given accordingly from the specific microscopic mechanism.
For the constant transition rates, our models retain the conventional
models describing the spreading dynamics with the probabilities $P_{l,i}(t)$
or the populations $n_{k,i}(t)$ and $N_{i}(t)$. The detailed derivation
is given in the supplementary materials \citep{supplementarymaterials}.

\begin{table*}[t]
\begin{tabular}{>{\centering}m{0.2\linewidth}>{\centering}p{0.3\linewidth}>{\centering}p{0.3\linewidth}}
\toprule 
 & SIS model & SIR model\tabularnewline
\midrule 
\rowcolor[HTML]{EFEFEF}

Node states & $0,1$ & $0,1,2$\tabularnewline
Rules & $1\overset{\alpha(\tau_{1})}{\longrightarrow}0$

$0+1\overset{\beta(\tau_{0},\tau_{1})}{\longrightarrow}1+1$ & $1\overset{\alpha(\tau_{1})}{\longrightarrow}2$

$0+1\overset{\beta(\tau_{0},\tau_{1})}{\longrightarrow}1+1$\tabularnewline
\rowcolor[HTML]{EFEFEF}

Transformation rates & \multicolumn{2}{c}{$\Gamma_{k,1}(\tau_{1},t)=\alpha(\tau_{1})$}\tabularnewline
\rowcolor[HTML]{EFEFEF} & \multicolumn{2}{c}{$\Gamma_{k,0}(\tau_{0},t)=k\int_{0}^{\infty}\beta(\tau_{0},\tau_{1})\sum_{k^{\prime}=1}^{\infty}P(k^{\prime}|k)\rho_{k^{\prime},1}(\tau_{1},t)d\tau_{1}$}\tabularnewline
Fluxes & $\Phi_{k,0}(t)=\int_{0}^{\infty}\alpha(\tau_{1})\rho_{k,1}(\tau_{1},t)d\tau_{1}$

$\Phi_{k,1}(t)=\int_{0}^{\infty}\Gamma_{k,0}(\tau_{0},t)\rho_{k,0}(\tau_{0},t)d\tau_{0}$ & $\Phi_{k,2}(t)=\int_{0}^{\infty}\alpha(\tau_{1})\rho_{k,1}(\tau_{1},t)d\tau_{1}$

$\Phi_{k,1}(t)=\int_{0}^{\infty}\Gamma_{k,0}(\tau_{0},t)\rho_{k,0}(\tau_{0},t)d\tau_{0}$\tabularnewline
\rowcolor[HTML]{EFEFEF}

Connecting conditions & $\rho_{k,0}(0,t)=\Phi_{k,0}(t)$

$\rho_{k,1}(0,t)=\Phi_{k,1}(t)$ & $\rho_{k,0}(0,t)=0$

$\rho_{k,1}(0,t)=\Phi_{k,1}(t)$

$\rho_{k,2}(0,t)=\Phi_{k,2}(t)$\tabularnewline
\bottomrule
\end{tabular}

\caption{The dictionary for constructing the duration-dependent SIS and SIR
model. \label{tab:Comparison-between-the}}
\end{table*}

As follows, we apply our spreading models to the epidemic spreading.
In Tab. \ref{tab:Comparison-between-the}, we list the dictionary
for constructing the duration-dependent SIS and SIR models with the
transformation rates, the fluxes and the connecting conditions in
the mesoscopic model. The two models are uniformly described by the
same partial differential equations with different coupling forms
of the connecting conditions.

The macroscopic model of spreading dynamics recovers to the standard
compartmental SIS model \citep{1927a,Bailey1975,Brauer2019} with
the constant recovery $\alpha$ and infection rate $\beta$, where
the susceptible and the infected populations satisfy $\dot{N}_{0}(t)=\alpha N_{1}(t)-\beta\left\langle k\right\rangle N_{0}(t)N_{1}(t)/N_{T}$
and $N_{1}(t)=N_{T}-N_{0}(t)$. In Ref. \citep{Du2020}, the effect
of the duration-dependent recovery rate $\alpha(\tau_{1})$ has been
studied in an extended compartmental model with the integro-differential
equations. In the supplementary materials \citep{supplementarymaterials},
we derive both the standard and the extended compartmental model from
the macroscopic model.

\textit{SIS model in a network.---}The current DCG approach is applied
to solve the spreading dynamics of the duration-dependent SIS model
on an uncorrelated network with the degree correlation $P(k^{\prime}|k)=k^{\prime}P(k^{\prime})/\left\langle k\right\rangle $
\citep{MarcBarthelemy2012}. The DCG approach enables us to obtain
the steady state with arbitrary duration-dependent recovery and infection
rates by solving a self-consistent equation.

In the duration-dependent SIS model, the DDFs $\rho_{k,0}(\tau_{0},t)$
and $\rho_{k,1}(\tau_{1},t)$ obey Eq. (\ref{eq:differentialequationofDDF1})
with the transformation rates and the connecting conditions listed
in Tab. \ref{tab:Comparison-between-the}. The epidemic spreading
is typically assessed by the fraction $r_{1}(t)=\left(\sum_{k=1}^{\infty}n_{k,1}(t)\right)/\left(\sum_{k=1}^{\infty}n_{k}\right)$
of the infected nodes. For the infection rate $\beta(\tau_{0},\tau_{1})$,
the dependence on the susceptible and the infection duration describes
the vulnerability of a susceptible node and the transmissibility of
an infected node, respectively. For simplicity, we assume the vulnerability
of the susceptible node does not rely on the susceptible duration
\citep{selfconsistentequation}. Namely, the spreading dynamics is
independent of the susceptible duration $\tau_{0}$, and the infection
rate only depends on the infection duration $\tau_{1}$ as $\beta(\tau_{0},\tau_{1})=\beta(\tau_{1})$.

On the uncorrelated network, the transformation rate of the contact
process is simplified as $\Gamma_{k,0}(t)=k\Theta(t)$ with

\begin{equation}
\Theta(t)=\sum_{k=1}^{\infty}\frac{kP(k)}{\left\langle k\right\rangle }\int_{0}^{\infty}\beta(\tau_{1})\rho_{k,1}(\tau_{1},t)d\tau_{1}.\label{eq:quantityTheta}
\end{equation}
For the steady state $\partial\rho_{k,i}(\tau_{i},t)/\partial t=0$
of Eq. (\ref{eq:differentialequationofDDF1}), the DDFs of the steady
state are solved as

\begin{equation}
\rho_{k,0}(\tau_{0})=\Phi_{k}\exp[-k\Theta\tau_{0}],\label{eq:steadystatefk0}
\end{equation}
and

\begin{equation}
\rho_{k,1}(\tau_{1})=\Phi_{k}\exp[-\int_{0}^{\tau_{1}}\alpha(\tau)d\tau].\label{eq:steadystatefk1}
\end{equation}
where $\Phi_{k}=n_{k}k\Theta/(1+k\Theta\bar{\tau}_{1})$ is the steady-state
flux with the average infection duration $\bar{\tau}_{1}=\int_{0}^{\infty}\exp[-\int_{0}^{\tau_{1}}\alpha(\tau)d\tau]d\tau_{1}$,
i.e., the average time to recover from the disease. It follows from
Eq. (\ref{eq:quantityTheta}) that

\begin{equation}
\Theta=\frac{\Upsilon\Theta}{\left\langle k\right\rangle }\sum_{k=1}^{\infty}\frac{k^{2}P(k)}{1+k\Theta\bar{\tau}_{1}},\label{eq:Thetaselfconsist}
\end{equation}
which is the self-consistent equation for the quantity $\Theta$ of
the steady state. Here, $\Upsilon$ is the refined spreading rate
for the duration-dependent SIS model as 
\begin{equation}
\Upsilon=\int_{0}^{\infty}\beta(\tau_{1})\exp[-\int_{0}^{\tau_{1}}\alpha(\tau)d\tau]d\tau_{1}.\label{eq:RefinedspreadingrateTheta}
\end{equation}
The steady-state fraction of the infected nodes is
\begin{equation}
r_{1}=\sum_{k=1}^{\infty}\frac{k\Theta\bar{\tau}_{1}}{1+k\Theta\bar{\tau}_{1}}P(k),\label{eq:defineinfectedfraction}
\end{equation}
which is determined by the refined spreading rate $\Upsilon$ via
the quantity $\Theta$ and the average infection duration $\bar{\tau}_{1}$
\citep{sincetheproduct}. The effect of network structure is explicitly
reflected via the degree distribution $P(k)$. For the constant recovery
and infection rates, the refined spreading rate $\Upsilon$ returns
to the effective spreading rate $\Upsilon=\beta/\alpha$ used in the
duration-independent SIS model \citep{Pastor-Satorras2001}.

The existence of the non-zero solution $\Theta$ requires the refined
spreading rate $\Upsilon$ to exceed a critical value $\Upsilon_{c}=\left\langle k\right\rangle /\left\langle k^{2}\right\rangle $,
which is defined as the epidemic threshold solely determined by the
network structure. When the refined spreading rate exceeds the epidemic
threshold $\Upsilon>\Upsilon_{c}$, the system reaches the epidemic
steady state with non-zero infected nodes. At the situation $\Upsilon<\Upsilon_{c}$,
the system reaches the disease-free steady state. A necessary condition
to ensure a disease-free steady state is $\left\langle k\right\rangle \leq1/\Upsilon$,
which implies the contacts of people need to be controlled according
to the spreading ability of the epidemic.

To validate the current coarse-grained model, we simulate the duration-dependent
SIS model in an uncorrelated scale-free network with the continuous-time
Monte Carlo method \citep{Mieghem2006,Li2012}. Details of the simulation
are illustrated in the supplementary materials \citep{supplementarymaterials}.
The uncorrelated scale-free network with $N_{T}=2500$ is generated
via the configuration model \citep{Catanzaro2005}. The degree sequence
$\{k_{l}\}$ is generated according to the degree distribution $P(k)=c/k^{3},$
where $k$ ranges from the minimal degree $k_{\mathrm{min}}=10$ to
the maximal degree $k_{\mathrm{max}}=50$ with the normalized constant
$c=1/(\sum_{k^{\prime}=k_{\mathrm{min}}}^{k_{\mathrm{max}}}1/k^{\prime3})$
of the degree distribution. The minimal degree $k_{\mathrm{min}}$
is chosen not so small to avoid large fluctuations of the infected
neighbors for low-degree nodes, since the mean-field approach assumes
the static PDF for the steady state without considering the fluctuations.
The maximal degree $k_{\mathrm{max}}$ fulfills the condition $k_{\mathrm{max}}\leq\sqrt{N_{T}}$
to ensure an uncorrelated network \citep{Catanzaro2005}. All nodes
are randomly linked respecting the assigned degrees without multiple
and self-connection.

We carry out the simulation with the Weibull distribution of the recovery
and the infection time obeying $\psi_{R}(\tau_{I})=a_{\alpha}/b_{\alpha}(\tau_{I}/b_{\alpha})^{a_{\alpha}-1}\exp[-(\tau_{I}/b_{\alpha})^{a_{\alpha}}]$
and $\psi_{I}(T_{I})=a_{\beta}/b_{\beta}(T_{I}/b_{\beta})^{a_{\beta}-1}\exp[-(T_{I}/b_{\beta})^{a_{\beta}}]$,
with the corresponding transition rates $\alpha(\tau)=a_{\alpha}/b_{\alpha}(\tau/b_{\alpha})^{a_{\alpha}-1}$
and $\beta(\tau)=a_{\beta}/b_{\beta}(\tau/b_{\beta})^{a_{\beta}-1}$.
In each simulation, the evolution is run for 500,000 events to reach
the steady state. The steady-state fraction $r_{1}$ is then obtained
as the average with 200,000 events.

\begin{figure}
\includegraphics[width=7cm]{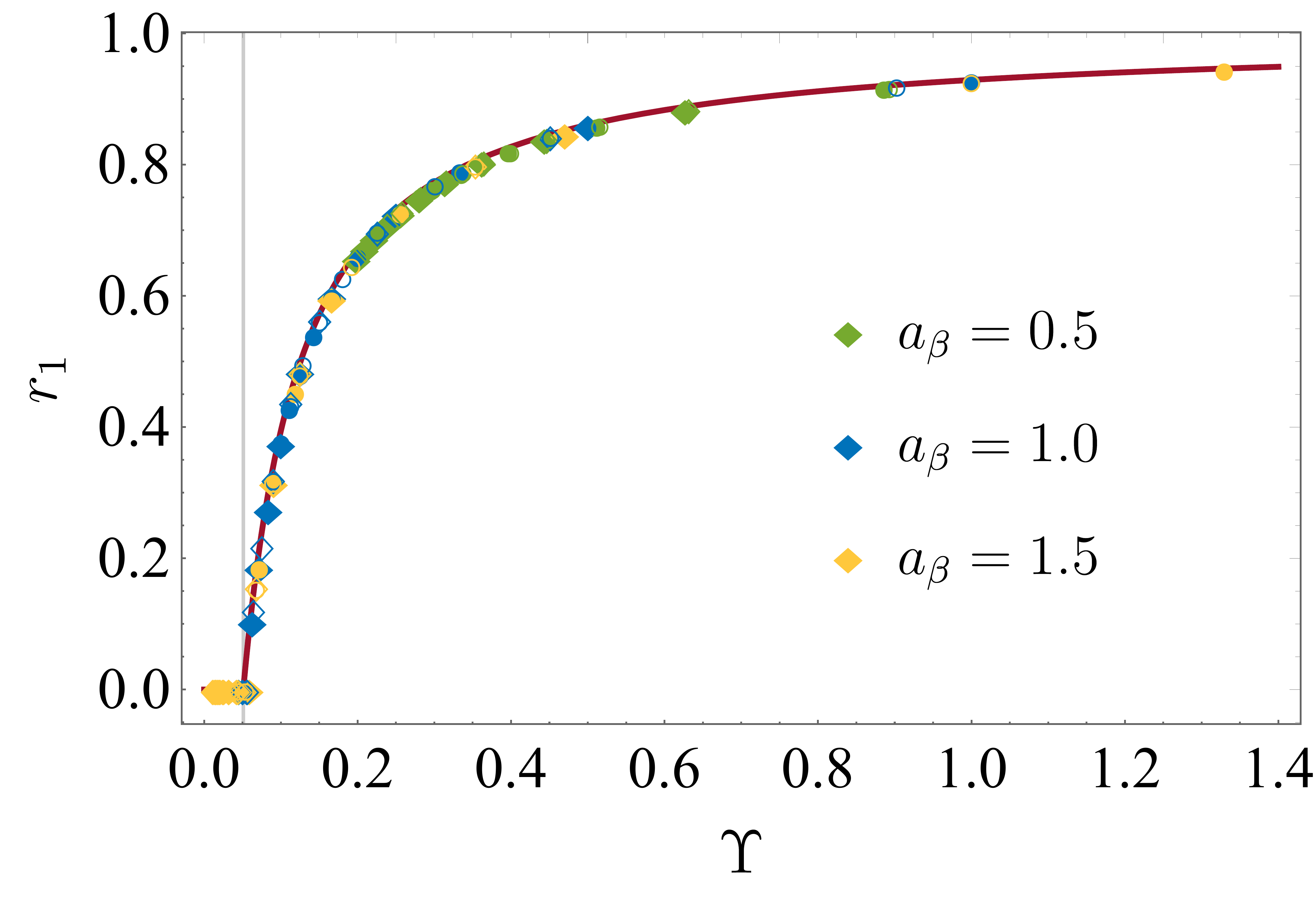}

\caption{(Color online) The steady-state fraction $r_{1}$ of the infected
nodes as the function of the refined spreading rate $\Upsilon$ in
the uncorrelated scale-free network. The solid curve is obtained by
the DCG approach according to Eq. (\ref{eq:defineinfectedfraction}).
The markers show the continuous-time Monte Carlo simulation results
with Weibull recovery and infection time, where the parameters are
set as $a_{\alpha}=1.0$ (filled), $1.5$ (empty), $b_{\alpha}=0.5$
(diamond), $1.0$ (circle) , $a_{\beta}=0.5,\,1.0,\,1.5$ (in different
colors), and $b_{\beta}$ ranging from $1.0$ to $10.0$ with the
interval $1.0$. The gray vertical line shows the epidemic threshold
$\Upsilon_{c}=0.051$ for the current finite-size scale-free network.
\label{fig:The-infected-ratio}}
\end{figure}

In Fig. \ref{fig:The-infected-ratio}, the steady-state fraction $r_{1}$
of the infected nodes is plotted as the function of the refined spreading
rate $\Upsilon$ for the DCG approach (solid curve) and the continuous-time
Monte Carlo simulation results (dots). In the simulation, the effects
of duration-dependent recovery and the infection rates are considered
with different sets of parameters. The agreement between the analytical
and the simulation results validates that the steady-state fraction
$r_{1}$ can be effectively described with the refined spreading rate
$\Upsilon$ by Eq. (\ref{eq:RefinedspreadingrateTheta}). The curve
shows that the existence of the epidemic threshold matches with the
theoretical prediction $\Upsilon_{c}=\left\langle k\right\rangle /\left\langle k^{2}\right\rangle =0.051$
(gray grid-line) . The current model shows the availability of the
refined spreading rate $\Upsilon$ for justifying the spreading ability
of an epidemic.

\textit{Conclusion.---}In this Letter, we generalize the mean-field
theories for the spreading dynamics with duration-dependent mechanism
by superseding the probability distribution of states with the PDF
of the duration, and show the hierarchical emergence of the widely-used
coarse-grained spreading models. The unified formalism enables us
to rebuild different epidemic models, e.g., the SIS and the SIR model.
Compared to Refs. \citep{Mieghem2013,Cator2013,Mieghem2019}, the
refined spreading rate $\Upsilon$ here is suitable for the duration-dependent
models as a coarse-grained parameter of the microscopic mechanism
details, and also suggests the duration-dependent SIS model can be
mapped to the standard one in the meaning of the steady states \citep{Starnini2017}.
With the refined spreading rate $\Upsilon$, the epidemic threshold
$\Upsilon_{c}=\left\langle k\right\rangle /\left\langle k^{2}\right\rangle $
is applicable for the duration-dependent SIS model to determine the
fate of the epidemic spreading.

Limited by the mean-field approach, the current formalism has neglected
correlations and fluctuations between nodes, and therefore cannot
accurately predict the critical point of the epidemic phase transition,
i.e., the epidemic threshold. In the standard SIS model, the correlations
and fluctuations affect the epidemic threshold through the mutual
reinfection of the high-degree nodes \citep{Goltsev2012,Boguna2013,Lee2013,Wei2020},
and was recently understood through the cumulative merging percolation
process \citep{Castellano2020}. It is still an open question to describe
such correlation effect in a duration-dependent model, which is beyond
the scope of the current work and worth for further investigation.
\begin{acknowledgments}
This work is supported by the NSFC (Grants No. 11534002), the NSAF
(Grant No. U1930403 and No. U1930402), and the National Basic Research
Program of China (Grants No. 2016YFA0301201).
\end{acknowledgments}

\bibliographystyle{apsrev4-1}
\bibliography{covidbib}

\begin{thebibliography}{62}%
\makeatletter
\providecommand \@ifxundefined [1]{%
 \@ifx{#1\undefined}
}%
\providecommand \@ifnum [1]{%
 \ifnum #1\expandafter \@firstoftwo
 \else \expandafter \@secondoftwo
 \fi
}%
\providecommand \@ifx [1]{%
 \ifx #1\expandafter \@firstoftwo
 \else \expandafter \@secondoftwo
 \fi
}%
\providecommand \natexlab [1]{#1}%
\providecommand \enquote  [1]{``#1''}%
\providecommand \bibnamefont  [1]{#1}%
\providecommand \bibfnamefont [1]{#1}%
\providecommand \citenamefont [1]{#1}%
\providecommand \href@noop [0]{\@secondoftwo}%
\providecommand \href [0]{\begingroup \@sanitize@url \@href}%
\providecommand \@href[1]{\@@startlink{#1}\@@href}%
\providecommand \@@href[1]{\endgroup#1\@@endlink}%
\providecommand \@sanitize@url [0]{\catcode `\\12\catcode `\$12\catcode
  `\&12\catcode `\#12\catcode `\^12\catcode `\_12\catcode `\%12\relax}%
\providecommand \@@startlink[1]{}%
\providecommand \@@endlink[0]{}%
\providecommand \url  [0]{\begingroup\@sanitize@url \@url }%
\providecommand \@url [1]{\endgroup\@href {#1}{\urlprefix }}%
\providecommand \urlprefix  [0]{URL }%
\providecommand \Eprint [0]{\href }%
\providecommand \doibase [0]{http://dx.doi.org/}%
\providecommand \selectlanguage [0]{\@gobble}%
\providecommand \bibinfo  [0]{\@secondoftwo}%
\providecommand \bibfield  [0]{\@secondoftwo}%
\providecommand \translation [1]{[#1]}%
\providecommand \BibitemOpen [0]{}%
\providecommand \bibitemStop [0]{}%
\providecommand \bibitemNoStop [0]{.\EOS\space}%
\providecommand \EOS [0]{\spacefactor3000\relax}%
\providecommand \BibitemShut  [1]{\csname bibitem#1\endcsname}%
\let\auto@bib@innerbib\@empty
\bibitem [{\citenamefont {Moore}\ and\ \citenamefont
  {Newman}(2000)}]{Moore2000}%
  \BibitemOpen
  \bibfield  {author} {\bibinfo {author} {\bibfnamefont {C.}~\bibnamefont
  {Moore}}\ and\ \bibinfo {author} {\bibfnamefont {M.~E.~J.}\ \bibnamefont
  {Newman}},\ }\href {\doibase 10.1103/physreve.61.5678} {\bibfield  {journal}
  {\bibinfo  {journal} {Phys. Rev. E}\ }\textbf {\bibinfo {volume} {61}},\
  \bibinfo {pages} {5678} (\bibinfo {year} {2000})}\BibitemShut {NoStop}%
\bibitem [{\citenamefont {Pastor-Satorras}\ and\ \citenamefont
  {Vespignani}(2001{\natexlab{a}})}]{Pastor-Satorras2001}%
  \BibitemOpen
  \bibfield  {author} {\bibinfo {author} {\bibfnamefont {R.}~\bibnamefont
  {Pastor-Satorras}}\ and\ \bibinfo {author} {\bibfnamefont {A.}~\bibnamefont
  {Vespignani}},\ }\href {\doibase 10.1103/physrevlett.86.3200} {\bibfield
  {journal} {\bibinfo  {journal} {Phys. Rev. Lett.}\ }\textbf {\bibinfo
  {volume} {86}},\ \bibinfo {pages} {3200} (\bibinfo {year}
  {2001}{\natexlab{a}})}\BibitemShut {NoStop}%
\bibitem [{\citenamefont {Pastor-Satorras}\ and\ \citenamefont
  {Vespignani}(2001{\natexlab{b}})}]{PastorSatorras2001}%
  \BibitemOpen
  \bibfield  {author} {\bibinfo {author} {\bibfnamefont {R.}~\bibnamefont
  {Pastor-Satorras}}\ and\ \bibinfo {author} {\bibfnamefont {A.}~\bibnamefont
  {Vespignani}},\ }\href {\doibase 10.1103/physreve.63.066117} {\bibfield
  {journal} {\bibinfo  {journal} {Phys. Rev. E}\ }\textbf {\bibinfo {volume}
  {63}},\ \bibinfo {pages} {066117} (\bibinfo {year}
  {2001}{\natexlab{b}})}\BibitemShut {NoStop}%
\bibitem [{\citenamefont {May}\ and\ \citenamefont {Lloyd}(2001)}]{May2001}%
  \BibitemOpen
  \bibfield  {author} {\bibinfo {author} {\bibfnamefont {R.~M.}\ \bibnamefont
  {May}}\ and\ \bibinfo {author} {\bibfnamefont {A.~L.}\ \bibnamefont
  {Lloyd}},\ }\href {\doibase 10.1103/physreve.64.066112} {\bibfield  {journal}
  {\bibinfo  {journal} {Phys. Rev. E}\ }\textbf {\bibinfo {volume} {64}},\
  \bibinfo {pages} {066112} (\bibinfo {year} {2001})}\BibitemShut {NoStop}%
\bibitem [{\citenamefont {Cai}\ \emph {et~al.}(2016)\citenamefont {Cai},
  \citenamefont {Wu}, \citenamefont {Chen}, \citenamefont {Holme},\ and\
  \citenamefont {Guan}}]{Cai2016}%
  \BibitemOpen
  \bibfield  {author} {\bibinfo {author} {\bibfnamefont {C.-R.}\ \bibnamefont
  {Cai}}, \bibinfo {author} {\bibfnamefont {Z.-X.}\ \bibnamefont {Wu}},
  \bibinfo {author} {\bibfnamefont {M.~Z.~Q.}\ \bibnamefont {Chen}}, \bibinfo
  {author} {\bibfnamefont {P.}~\bibnamefont {Holme}}, \ and\ \bibinfo {author}
  {\bibfnamefont {J.-Y.}\ \bibnamefont {Guan}},\ }\href {\doibase
  10.1103/physrevlett.116.258301} {\bibfield  {journal} {\bibinfo  {journal}
  {Phys. Rev. Lett.}\ }\textbf {\bibinfo {volume} {116}},\ \bibinfo {pages}
  {258301} (\bibinfo {year} {2016})}\BibitemShut {NoStop}%
\bibitem [{\citenamefont {Hindes}\ and\ \citenamefont
  {Schwartz}(2016)}]{Hindes2016}%
  \BibitemOpen
  \bibfield  {author} {\bibinfo {author} {\bibfnamefont {J.}~\bibnamefont
  {Hindes}}\ and\ \bibinfo {author} {\bibfnamefont {I.~B.}\ \bibnamefont
  {Schwartz}},\ }\href {\doibase 10.1103/physrevlett.117.028302} {\bibfield
  {journal} {\bibinfo  {journal} {Phys. Rev. Lett.}\ }\textbf {\bibinfo
  {volume} {117}},\ \bibinfo {pages} {028302} (\bibinfo {year}
  {2016})}\BibitemShut {NoStop}%
\bibitem [{\citenamefont {Goffman}\ and\ \citenamefont
  {Newill}(1964)}]{GOFFMAN1964}%
  \BibitemOpen
  \bibfield  {author} {\bibinfo {author} {\bibfnamefont {W.}~\bibnamefont
  {Goffman}}\ and\ \bibinfo {author} {\bibfnamefont {V.~A.}\ \bibnamefont
  {Newill}},\ }\href {\doibase 10.1038/204225a0} {\bibfield  {journal}
  {\bibinfo  {journal} {Nature}\ }\textbf {\bibinfo {volume} {204}},\ \bibinfo
  {pages} {225} (\bibinfo {year} {1964})}\BibitemShut {NoStop}%
\bibitem [{\citenamefont {Daley}\ and\ \citenamefont
  {Kendall}(1964)}]{DALEY1964}%
  \BibitemOpen
  \bibfield  {author} {\bibinfo {author} {\bibfnamefont {D.~J.}\ \bibnamefont
  {Daley}}\ and\ \bibinfo {author} {\bibfnamefont {D.~G.}\ \bibnamefont
  {Kendall}},\ }\href {\doibase 10.1038/2041118a0} {\bibfield  {journal}
  {\bibinfo  {journal} {Nature}\ }\textbf {\bibinfo {volume} {204}},\ \bibinfo
  {pages} {1118} (\bibinfo {year} {1964})}\BibitemShut {NoStop}%
\bibitem [{\citenamefont {Moreno}\ \emph {et~al.}(2004)\citenamefont {Moreno},
  \citenamefont {Nekovee},\ and\ \citenamefont {Pacheco}}]{Moreno2004}%
  \BibitemOpen
  \bibfield  {author} {\bibinfo {author} {\bibfnamefont {Y.}~\bibnamefont
  {Moreno}}, \bibinfo {author} {\bibfnamefont {M.}~\bibnamefont {Nekovee}}, \
  and\ \bibinfo {author} {\bibfnamefont {A.~F.}\ \bibnamefont {Pacheco}},\
  }\href {\doibase 10.1103/physreve.69.066130} {\bibfield  {journal} {\bibinfo
  {journal} {Phys. Rev. E}\ }\textbf {\bibinfo {volume} {69}},\ \bibinfo
  {pages} {066130} (\bibinfo {year} {2004})}\BibitemShut {NoStop}%
\bibitem [{\citenamefont {Nematzadeh}\ \emph {et~al.}(2014)\citenamefont
  {Nematzadeh}, \citenamefont {Ferrara}, \citenamefont {Flammini},\ and\
  \citenamefont {Ahn}}]{Nematzadeh2014}%
  \BibitemOpen
  \bibfield  {author} {\bibinfo {author} {\bibfnamefont {A.}~\bibnamefont
  {Nematzadeh}}, \bibinfo {author} {\bibfnamefont {E.}~\bibnamefont {Ferrara}},
  \bibinfo {author} {\bibfnamefont {A.}~\bibnamefont {Flammini}}, \ and\
  \bibinfo {author} {\bibfnamefont {Y.-Y.}\ \bibnamefont {Ahn}},\ }\href
  {\doibase 10.1103/physrevlett.113.088701} {\bibfield  {journal} {\bibinfo
  {journal} {Phys. Rev. Lett.}\ }\textbf {\bibinfo {volume} {113}},\ \bibinfo
  {pages} {088701} (\bibinfo {year} {2014})}\BibitemShut {NoStop}%
\bibitem [{\citenamefont {Gleeson}\ \emph {et~al.}(2016)\citenamefont
  {Gleeson}, \citenamefont {O'Sullivan}, \citenamefont {Ba{\~{n}}os},\ and\
  \citenamefont {Moreno}}]{Gleeson2016}%
  \BibitemOpen
  \bibfield  {author} {\bibinfo {author} {\bibfnamefont {J.~P.}\ \bibnamefont
  {Gleeson}}, \bibinfo {author} {\bibfnamefont {K.~P.}\ \bibnamefont
  {O'Sullivan}}, \bibinfo {author} {\bibfnamefont {R.~A.}\ \bibnamefont
  {Ba{\~{n}}os}}, \ and\ \bibinfo {author} {\bibfnamefont {Y.}~\bibnamefont
  {Moreno}},\ }\href {\doibase 10.1103/physrevx.6.021019} {\bibfield  {journal}
  {\bibinfo  {journal} {Phys. Rev. X}\ }\textbf {\bibinfo {volume} {6}},\
  \bibinfo {pages} {021019} (\bibinfo {year} {2016})}\BibitemShut {NoStop}%
\bibitem [{\citenamefont {K.~Sznajd-Weron}(2000)}]{SZNAJD-WERON2000}%
  \BibitemOpen
  \bibfield  {author} {\bibinfo {author} {\bibfnamefont {J.~S.}\ \bibnamefont
  {K.~Sznajd-Weron}},\ }\href {\doibase 10.1142/s0129183100000936} {\bibfield
  {journal} {\bibinfo  {journal} {Int. J. Mod. Phys. C}\ }\textbf {\bibinfo
  {volume} {11}},\ \bibinfo {pages} {1157} (\bibinfo {year}
  {2000})}\BibitemShut {NoStop}%
\bibitem [{\citenamefont {Dornic}\ \emph {et~al.}(2001)\citenamefont {Dornic},
  \citenamefont {Chat{\'{e}}}, \citenamefont {Chave},\ and\ \citenamefont
  {Hinrichsen}}]{Dornic2001}%
  \BibitemOpen
  \bibfield  {author} {\bibinfo {author} {\bibfnamefont {I.}~\bibnamefont
  {Dornic}}, \bibinfo {author} {\bibfnamefont {H.}~\bibnamefont {Chat{\'{e}}}},
  \bibinfo {author} {\bibfnamefont {J.}~\bibnamefont {Chave}}, \ and\ \bibinfo
  {author} {\bibfnamefont {H.}~\bibnamefont {Hinrichsen}},\ }\href {\doibase
  10.1103/physrevlett.87.045701} {\bibfield  {journal} {\bibinfo  {journal}
  {Phys. Rev. Lett.}\ }\textbf {\bibinfo {volume} {87}},\ \bibinfo {pages}
  {045701} (\bibinfo {year} {2001})}\BibitemShut {NoStop}%
\bibitem [{\citenamefont {Krapivsky}\ and\ \citenamefont
  {Redner}(2003)}]{Krapivsky2003}%
  \BibitemOpen
  \bibfield  {author} {\bibinfo {author} {\bibfnamefont {P.~L.}\ \bibnamefont
  {Krapivsky}}\ and\ \bibinfo {author} {\bibfnamefont {S.}~\bibnamefont
  {Redner}},\ }\href {\doibase 10.1103/physrevlett.90.238701} {\bibfield
  {journal} {\bibinfo  {journal} {Phys. Rev. Lett.}\ }\textbf {\bibinfo
  {volume} {90}},\ \bibinfo {pages} {238701} (\bibinfo {year}
  {2003})}\BibitemShut {NoStop}%
\bibitem [{\citenamefont {Fern{\'{a}}ndez-Gracia}\ \emph
  {et~al.}(2014)\citenamefont {Fern{\'{a}}ndez-Gracia}, \citenamefont
  {Suchecki}, \citenamefont {Ramasco}, \citenamefont {SanMiguel},\ and\
  \citenamefont {Egu{\'{\i}}luz}}]{Fernandez-Gracia2014}%
  \BibitemOpen
  \bibfield  {author} {\bibinfo {author} {\bibfnamefont {J.}~\bibnamefont
  {Fern{\'{a}}ndez-Gracia}}, \bibinfo {author} {\bibfnamefont {K.}~\bibnamefont
  {Suchecki}}, \bibinfo {author} {\bibfnamefont {J.~J.}\ \bibnamefont
  {Ramasco}}, \bibinfo {author} {\bibfnamefont {M.}~\bibnamefont {SanMiguel}},
  \ and\ \bibinfo {author} {\bibfnamefont {V.~M.}\ \bibnamefont
  {Egu{\'{\i}}luz}},\ }\href {\doibase 10.1103/physrevlett.112.158701}
  {\bibfield  {journal} {\bibinfo  {journal} {Phys. Rev. Lett.}\ }\textbf
  {\bibinfo {volume} {112}},\ \bibinfo {pages} {158701} (\bibinfo {year}
  {2014})}\BibitemShut {NoStop}%
\bibitem [{\citenamefont {Albert}\ and\ \citenamefont
  {Barab{\'{a}}si}(2002)}]{Albert2002}%
  \BibitemOpen
  \bibfield  {author} {\bibinfo {author} {\bibfnamefont {R.}~\bibnamefont
  {Albert}}\ and\ \bibinfo {author} {\bibfnamefont {A.-L.}\ \bibnamefont
  {Barab{\'{a}}si}},\ }\href {\doibase 10.1103/revmodphys.74.47} {\bibfield
  {journal} {\bibinfo  {journal} {Rev. Mod. Phys.}\ }\textbf {\bibinfo {volume}
  {74}},\ \bibinfo {pages} {47} (\bibinfo {year} {2002})}\BibitemShut {NoStop}%
\bibitem [{\citenamefont {Barrat}\ \emph {et~al.}(2012)\citenamefont {Barrat},
  \citenamefont {Barthelemy},\ and\ \citenamefont
  {Vespignani}}]{MarcBarthelemy2012}%
  \BibitemOpen
  \bibfield  {author} {\bibinfo {author} {\bibfnamefont {A.}~\bibnamefont
  {Barrat}}, \bibinfo {author} {\bibfnamefont {M.}~\bibnamefont {Barthelemy}},
  \ and\ \bibinfo {author} {\bibfnamefont {A.}~\bibnamefont {Vespignani}},\
  }\href
  {https://www.ebook.de/de/product/7871347/marc_barthelemy_alessandro_vespignani_dynamical_processes_on_complex_networks.html}
  {\emph {\bibinfo {title} {Dynamical Processes on Complex Networks}}}\
  (\bibinfo  {publisher} {Cambridge University Press},\ \bibinfo {year}
  {2012})\BibitemShut {NoStop}%
\bibitem [{\citenamefont {Chakrabarti}\ \emph {et~al.}(2008)\citenamefont
  {Chakrabarti}, \citenamefont {Wang}, \citenamefont {Wang}, \citenamefont
  {Leskovec},\ and\ \citenamefont {Faloutsos}}]{Chakrabarti2008}%
  \BibitemOpen
  \bibfield  {author} {\bibinfo {author} {\bibfnamefont {D.}~\bibnamefont
  {Chakrabarti}}, \bibinfo {author} {\bibfnamefont {Y.}~\bibnamefont {Wang}},
  \bibinfo {author} {\bibfnamefont {C.}~\bibnamefont {Wang}}, \bibinfo {author}
  {\bibfnamefont {J.}~\bibnamefont {Leskovec}}, \ and\ \bibinfo {author}
  {\bibfnamefont {C.}~\bibnamefont {Faloutsos}},\ }\href {\doibase
  10.1145/1284680.1284681} {\bibfield  {journal} {\bibinfo  {journal} {ACM
  Trans. Inf. Syst. Secur.}\ }\textbf {\bibinfo {volume} {10}},\ \bibinfo
  {pages} {1} (\bibinfo {year} {2008})}\BibitemShut {NoStop}%
\bibitem [{\citenamefont {Mieghem}\ \emph {et~al.}(2009)\citenamefont
  {Mieghem}, \citenamefont {Omic},\ and\ \citenamefont {Kooij}}]{Mieghem2009}%
  \BibitemOpen
  \bibfield  {author} {\bibinfo {author} {\bibfnamefont {P.~V.}\ \bibnamefont
  {Mieghem}}, \bibinfo {author} {\bibfnamefont {J.}~\bibnamefont {Omic}}, \
  and\ \bibinfo {author} {\bibfnamefont {R.}~\bibnamefont {Kooij}},\ }\href
  {\doibase 10.1109/tnet.2008.925623} {\bibfield  {journal} {\bibinfo
  {journal} {IEEE ACM Trans. Netw.}\ }\textbf {\bibinfo {volume} {17}},\
  \bibinfo {pages} {1} (\bibinfo {year} {2009})}\BibitemShut {NoStop}%
\bibitem [{\citenamefont {G{\'{o}}mez}\ \emph {et~al.}(2010)\citenamefont
  {G{\'{o}}mez}, \citenamefont {Arenas}, \citenamefont {Borge-Holthoefer},
  \citenamefont {Meloni},\ and\ \citenamefont {Moreno}}]{Gomez2010}%
  \BibitemOpen
  \bibfield  {author} {\bibinfo {author} {\bibfnamefont {S.}~\bibnamefont
  {G{\'{o}}mez}}, \bibinfo {author} {\bibfnamefont {A.}~\bibnamefont {Arenas}},
  \bibinfo {author} {\bibfnamefont {J.}~\bibnamefont {Borge-Holthoefer}},
  \bibinfo {author} {\bibfnamefont {S.}~\bibnamefont {Meloni}}, \ and\ \bibinfo
  {author} {\bibfnamefont {Y.}~\bibnamefont {Moreno}},\ }\href {\doibase
  10.1209/0295-5075/89/38009} {\bibfield  {journal} {\bibinfo  {journal}
  {Europhys. Lett.}\ }\textbf {\bibinfo {volume} {89}},\ \bibinfo {pages}
  {38009} (\bibinfo {year} {2010})}\BibitemShut {NoStop}%
\bibitem [{\citenamefont {Castellano}\ and\ \citenamefont
  {Pastor-Satorras}(2010)}]{Castellano2010}%
  \BibitemOpen
  \bibfield  {author} {\bibinfo {author} {\bibfnamefont {C.}~\bibnamefont
  {Castellano}}\ and\ \bibinfo {author} {\bibfnamefont {R.}~\bibnamefont
  {Pastor-Satorras}},\ }\href {\doibase 10.1103/physrevlett.105.218701}
  {\bibfield  {journal} {\bibinfo  {journal} {Phys. Rev. Lett.}\ }\textbf
  {\bibinfo {volume} {105}},\ \bibinfo {pages} {218701} (\bibinfo {year}
  {2010})}\BibitemShut {NoStop}%
\bibitem [{\citenamefont {Ferreira}\ \emph {et~al.}(2012)\citenamefont
  {Ferreira}, \citenamefont {Castellano},\ and\ \citenamefont
  {Pastor-Satorras}}]{Ferreira2012}%
  \BibitemOpen
  \bibfield  {author} {\bibinfo {author} {\bibfnamefont {S.~C.}\ \bibnamefont
  {Ferreira}}, \bibinfo {author} {\bibfnamefont {C.}~\bibnamefont
  {Castellano}}, \ and\ \bibinfo {author} {\bibfnamefont {R.}~\bibnamefont
  {Pastor-Satorras}},\ }\href {\doibase 10.1103/physreve.86.041125} {\bibfield
  {journal} {\bibinfo  {journal} {Phys. Rev. E}\ }\textbf {\bibinfo {volume}
  {86}},\ \bibinfo {pages} {041125} (\bibinfo {year} {2012})}\BibitemShut
  {NoStop}%
\bibitem [{\citenamefont {Li}\ \emph {et~al.}(2012)\citenamefont {Li},
  \citenamefont {van~de Bovenkamp},\ and\ \citenamefont {Mieghem}}]{Li2012}%
  \BibitemOpen
  \bibfield  {author} {\bibinfo {author} {\bibfnamefont {C.}~\bibnamefont
  {Li}}, \bibinfo {author} {\bibfnamefont {R.}~\bibnamefont {van~de
  Bovenkamp}}, \ and\ \bibinfo {author} {\bibfnamefont {P.~V.}\ \bibnamefont
  {Mieghem}},\ }\href {\doibase 10.1103/physreve.86.026116} {\bibfield
  {journal} {\bibinfo  {journal} {Phys. Rev. E}\ }\textbf {\bibinfo {volume}
  {86}},\ \bibinfo {pages} {026116} (\bibinfo {year} {2012})}\BibitemShut
  {NoStop}%
\bibitem [{\citenamefont {Goltsev}\ \emph {et~al.}(2012)\citenamefont
  {Goltsev}, \citenamefont {Dorogovtsev}, \citenamefont {Oliveira},\ and\
  \citenamefont {Mendes}}]{Goltsev2012}%
  \BibitemOpen
  \bibfield  {author} {\bibinfo {author} {\bibfnamefont {A.~V.}\ \bibnamefont
  {Goltsev}}, \bibinfo {author} {\bibfnamefont {S.~N.}\ \bibnamefont
  {Dorogovtsev}}, \bibinfo {author} {\bibfnamefont {J.~G.}\ \bibnamefont
  {Oliveira}}, \ and\ \bibinfo {author} {\bibfnamefont {J.~F.~F.}\ \bibnamefont
  {Mendes}},\ }\href {\doibase 10.1103/physrevlett.109.128702} {\bibfield
  {journal} {\bibinfo  {journal} {Phys. Rev. Lett.}\ }\textbf {\bibinfo
  {volume} {109}},\ \bibinfo {pages} {128702} (\bibinfo {year}
  {2012})}\BibitemShut {NoStop}%
\bibitem [{\citenamefont {Lee}\ \emph {et~al.}(2013)\citenamefont {Lee},
  \citenamefont {Shim},\ and\ \citenamefont {Noh}}]{Lee2013}%
  \BibitemOpen
  \bibfield  {author} {\bibinfo {author} {\bibfnamefont {H.~K.}\ \bibnamefont
  {Lee}}, \bibinfo {author} {\bibfnamefont {P.-S.}\ \bibnamefont {Shim}}, \
  and\ \bibinfo {author} {\bibfnamefont {J.~D.}\ \bibnamefont {Noh}},\ }\href
  {\doibase 10.1103/physreve.87.062812} {\bibfield  {journal} {\bibinfo
  {journal} {Phys. Rev. E}\ }\textbf {\bibinfo {volume} {87}},\ \bibinfo
  {pages} {062812} (\bibinfo {year} {2013})}\BibitemShut {NoStop}%
\bibitem [{\citenamefont {Bogu{\~{n}}{\'{a}}}\ \emph
  {et~al.}(2013)\citenamefont {Bogu{\~{n}}{\'{a}}}, \citenamefont
  {Castellano},\ and\ \citenamefont {Pastor-Satorras}}]{Boguna2013}%
  \BibitemOpen
  \bibfield  {author} {\bibinfo {author} {\bibfnamefont {M.}~\bibnamefont
  {Bogu{\~{n}}{\'{a}}}}, \bibinfo {author} {\bibfnamefont {C.}~\bibnamefont
  {Castellano}}, \ and\ \bibinfo {author} {\bibfnamefont {R.}~\bibnamefont
  {Pastor-Satorras}},\ }\href {\doibase 10.1103/physrevlett.111.068701}
  {\bibfield  {journal} {\bibinfo  {journal} {Phys. Rev. Lett.}\ }\textbf
  {\bibinfo {volume} {111}},\ \bibinfo {pages} {068701} (\bibinfo {year}
  {2013})}\BibitemShut {NoStop}%
\bibitem [{\citenamefont {Castellano}\ and\ \citenamefont
  {Pastor-Satorras}(2017)}]{Castellano2017}%
  \BibitemOpen
  \bibfield  {author} {\bibinfo {author} {\bibfnamefont {C.}~\bibnamefont
  {Castellano}}\ and\ \bibinfo {author} {\bibfnamefont {R.}~\bibnamefont
  {Pastor-Satorras}},\ }\href {\doibase 10.1103/physrevx.7.041024} {\bibfield
  {journal} {\bibinfo  {journal} {Phys. Rev. X}\ }\textbf {\bibinfo {volume}
  {7}},\ \bibinfo {pages} {041024} (\bibinfo {year} {2017})}\BibitemShut
  {NoStop}%
\bibitem [{\citenamefont {Parshani}\ \emph {et~al.}(2010)\citenamefont
  {Parshani}, \citenamefont {Carmi},\ and\ \citenamefont
  {Havlin}}]{Parshani2010}%
  \BibitemOpen
  \bibfield  {author} {\bibinfo {author} {\bibfnamefont {R.}~\bibnamefont
  {Parshani}}, \bibinfo {author} {\bibfnamefont {S.}~\bibnamefont {Carmi}}, \
  and\ \bibinfo {author} {\bibfnamefont {S.}~\bibnamefont {Havlin}},\ }\href
  {\doibase 10.1103/physrevlett.104.258701} {\bibfield  {journal} {\bibinfo
  {journal} {Phys. Rev. Lett.}\ }\textbf {\bibinfo {volume} {104}},\ \bibinfo
  {pages} {258701} (\bibinfo {year} {2010})}\BibitemShut {NoStop}%
\bibitem [{\citenamefont {Wei}\ and\ \citenamefont {Wang}(2020)}]{Wei2020}%
  \BibitemOpen
  \bibfield  {author} {\bibinfo {author} {\bibfnamefont {Z.-W.}\ \bibnamefont
  {Wei}}\ and\ \bibinfo {author} {\bibfnamefont {B.-H.}\ \bibnamefont {Wang}},\
  }\href {\doibase 10.1103/physreve.101.042310} {\bibfield  {journal} {\bibinfo
   {journal} {Phys. Rev. E}\ }\textbf {\bibinfo {volume} {101}},\ \bibinfo
  {pages} {042310} (\bibinfo {year} {2020})}\BibitemShut {NoStop}%
\bibitem [{\citenamefont {Dorogovtsev}\ \emph {et~al.}(2008)\citenamefont
  {Dorogovtsev}, \citenamefont {Goltsev},\ and\ \citenamefont
  {Mendes}}]{Dorogovtsev2008}%
  \BibitemOpen
  \bibfield  {author} {\bibinfo {author} {\bibfnamefont {S.~N.}\ \bibnamefont
  {Dorogovtsev}}, \bibinfo {author} {\bibfnamefont {A.~V.}\ \bibnamefont
  {Goltsev}}, \ and\ \bibinfo {author} {\bibfnamefont {J.~F.~F.}\ \bibnamefont
  {Mendes}},\ }\href {\doibase 10.1103/revmodphys.80.1275} {\bibfield
  {journal} {\bibinfo  {journal} {Rev. Mod. Phys.}\ }\textbf {\bibinfo {volume}
  {80}},\ \bibinfo {pages} {1275} (\bibinfo {year} {2008})}\BibitemShut
  {NoStop}%
\bibitem [{\citenamefont {Castellano}\ \emph {et~al.}(2009)\citenamefont
  {Castellano}, \citenamefont {Fortunato},\ and\ \citenamefont
  {Loreto}}]{Castellano2009}%
  \BibitemOpen
  \bibfield  {author} {\bibinfo {author} {\bibfnamefont {C.}~\bibnamefont
  {Castellano}}, \bibinfo {author} {\bibfnamefont {S.}~\bibnamefont
  {Fortunato}}, \ and\ \bibinfo {author} {\bibfnamefont {V.}~\bibnamefont
  {Loreto}},\ }\href {\doibase 10.1103/revmodphys.81.591} {\bibfield  {journal}
  {\bibinfo  {journal} {Rev. Mod. Phys.}\ }\textbf {\bibinfo {volume} {81}},\
  \bibinfo {pages} {591} (\bibinfo {year} {2009})}\BibitemShut {NoStop}%
\bibitem [{\citenamefont {Pastor-Satorras}\ \emph {et~al.}(2015)\citenamefont
  {Pastor-Satorras}, \citenamefont {Castellano}, \citenamefont {Mieghem},\ and\
  \citenamefont {Vespignani}}]{PastorSatorras2015}%
  \BibitemOpen
  \bibfield  {author} {\bibinfo {author} {\bibfnamefont {R.}~\bibnamefont
  {Pastor-Satorras}}, \bibinfo {author} {\bibfnamefont {C.}~\bibnamefont
  {Castellano}}, \bibinfo {author} {\bibfnamefont {P.~V.}\ \bibnamefont
  {Mieghem}}, \ and\ \bibinfo {author} {\bibfnamefont {A.}~\bibnamefont
  {Vespignani}},\ }\href {\doibase 10.1103/revmodphys.87.925} {\bibfield
  {journal} {\bibinfo  {journal} {Rev. Mod. Phys.}\ }\textbf {\bibinfo {volume}
  {87}},\ \bibinfo {pages} {925} (\bibinfo {year} {2015})}\BibitemShut
  {NoStop}%
\bibitem [{\citenamefont {Hoppensteadt}(1974)}]{Hoppensteadt1974}%
  \BibitemOpen
  \bibfield  {author} {\bibinfo {author} {\bibfnamefont {F.}~\bibnamefont
  {Hoppensteadt}},\ }\href {\doibase 10.1016/0016-0032(74)90037-4} {\bibfield
  {journal} {\bibinfo  {journal} {J. Franklin Inst.}\ }\textbf {\bibinfo
  {volume} {297}},\ \bibinfo {pages} {325} (\bibinfo {year}
  {1974})}\BibitemShut {NoStop}%
\bibitem [{\citenamefont {Feng}\ \emph {et~al.}(2007)\citenamefont {Feng},
  \citenamefont {Xu},\ and\ \citenamefont {Zhao}}]{Feng2007}%
  \BibitemOpen
  \bibfield  {author} {\bibinfo {author} {\bibfnamefont {Z.}~\bibnamefont
  {Feng}}, \bibinfo {author} {\bibfnamefont {D.}~\bibnamefont {Xu}}, \ and\
  \bibinfo {author} {\bibfnamefont {H.}~\bibnamefont {Zhao}},\ }\href {\doibase
  10.1007/s11538-006-9174-9} {\bibfield  {journal} {\bibinfo  {journal} {Bull.
  Math. Biol.}\ }\textbf {\bibinfo {volume} {69}},\ \bibinfo {pages} {1511}
  (\bibinfo {year} {2007})}\BibitemShut {NoStop}%
\bibitem [{\citenamefont {Magal}\ \emph {et~al.}(2010)\citenamefont {Magal},
  \citenamefont {McCluskey},\ and\ \citenamefont {Webb}}]{Magal2010}%
  \BibitemOpen
  \bibfield  {author} {\bibinfo {author} {\bibfnamefont {P.}~\bibnamefont
  {Magal}}, \bibinfo {author} {\bibfnamefont {C.}~\bibnamefont {McCluskey}}, \
  and\ \bibinfo {author} {\bibfnamefont {G.}~\bibnamefont {Webb}},\ }\href
  {\doibase 10.1080/00036810903208122} {\bibfield  {journal} {\bibinfo
  {journal} {Appl. Anal.}\ }\textbf {\bibinfo {volume} {89}},\ \bibinfo {pages}
  {1109} (\bibinfo {year} {2010})}\BibitemShut {NoStop}%
\bibitem [{\citenamefont {Liu}\ \emph {et~al.}(2015)\citenamefont {Liu},
  \citenamefont {Wang},\ and\ \citenamefont {Liu}}]{Liu2015}%
  \BibitemOpen
  \bibfield  {author} {\bibinfo {author} {\bibfnamefont {L.}~\bibnamefont
  {Liu}}, \bibinfo {author} {\bibfnamefont {J.}~\bibnamefont {Wang}}, \ and\
  \bibinfo {author} {\bibfnamefont {X.}~\bibnamefont {Liu}},\ }\href {\doibase
  10.1016/j.nonrwa.2015.01.001} {\bibfield  {journal} {\bibinfo  {journal}
  {Nonlinear Anal. Real World Appl.}\ }\textbf {\bibinfo {volume} {24}},\
  \bibinfo {pages} {18} (\bibinfo {year} {2015})}\BibitemShut {NoStop}%
\bibitem [{\citenamefont {Wang}\ \emph {et~al.}(2016)\citenamefont {Wang},
  \citenamefont {Liu},\ and\ \citenamefont {Zhang}}]{Wang2016}%
  \BibitemOpen
  \bibfield  {author} {\bibinfo {author} {\bibfnamefont {L.}~\bibnamefont
  {Wang}}, \bibinfo {author} {\bibfnamefont {Z.}~\bibnamefont {Liu}}, \ and\
  \bibinfo {author} {\bibfnamefont {X.}~\bibnamefont {Zhang}},\ }\href
  {\doibase 10.1016/j.nonrwa.2016.04.009} {\bibfield  {journal} {\bibinfo
  {journal} {Nonlinear Anal. Real World Appl.}\ }\textbf {\bibinfo {volume}
  {32}},\ \bibinfo {pages} {136} (\bibinfo {year} {2016})}\BibitemShut
  {NoStop}%
\bibitem [{\citenamefont {Mieghem}\ and\ \citenamefont {van~de
  Bovenkamp}(2013)}]{Mieghem2013}%
  \BibitemOpen
  \bibfield  {author} {\bibinfo {author} {\bibfnamefont {P.~V.}\ \bibnamefont
  {Mieghem}}\ and\ \bibinfo {author} {\bibfnamefont {R.}~\bibnamefont {van~de
  Bovenkamp}},\ }\href {\doibase 10.1103/physrevlett.110.108701} {\bibfield
  {journal} {\bibinfo  {journal} {Phys. Rev. Lett.}\ }\textbf {\bibinfo
  {volume} {110}},\ \bibinfo {pages} {108701} (\bibinfo {year}
  {2013})}\BibitemShut {NoStop}%
\bibitem [{\citenamefont {Cator}\ \emph {et~al.}(2013)\citenamefont {Cator},
  \citenamefont {van~de Bovenkamp},\ and\ \citenamefont {Mieghem}}]{Cator2013}%
  \BibitemOpen
  \bibfield  {author} {\bibinfo {author} {\bibfnamefont {E.}~\bibnamefont
  {Cator}}, \bibinfo {author} {\bibfnamefont {R.}~\bibnamefont {van~de
  Bovenkamp}}, \ and\ \bibinfo {author} {\bibfnamefont {P.~V.}\ \bibnamefont
  {Mieghem}},\ }\href {\doibase 10.1103/physreve.87.062816} {\bibfield
  {journal} {\bibinfo  {journal} {Phys. Rev. E}\ }\textbf {\bibinfo {volume}
  {87}},\ \bibinfo {pages} {062816} (\bibinfo {year} {2013})}\BibitemShut
  {NoStop}%
\bibitem [{\citenamefont {Yang}\ \emph {et~al.}(2016)\citenamefont {Yang},
  \citenamefont {Chen},\ and\ \citenamefont {Xu}}]{Yang2016}%
  \BibitemOpen
  \bibfield  {author} {\bibinfo {author} {\bibfnamefont {J.}~\bibnamefont
  {Yang}}, \bibinfo {author} {\bibfnamefont {Y.}~\bibnamefont {Chen}}, \ and\
  \bibinfo {author} {\bibfnamefont {F.}~\bibnamefont {Xu}},\ }\href {\doibase
  10.1007/s00285-016-0991-7} {\bibfield  {journal} {\bibinfo  {journal} {J.
  Math. Biol.}\ }\textbf {\bibinfo {volume} {73}},\ \bibinfo {pages} {1227}
  (\bibinfo {year} {2016})}\BibitemShut {NoStop}%
\bibitem [{\citenamefont {Chen}\ \emph {et~al.}(2018)\citenamefont {Chen},
  \citenamefont {Small}, \citenamefont {Tao},\ and\ \citenamefont
  {Fu}}]{Chen2018}%
  \BibitemOpen
  \bibfield  {author} {\bibinfo {author} {\bibfnamefont {S.}~\bibnamefont
  {Chen}}, \bibinfo {author} {\bibfnamefont {M.}~\bibnamefont {Small}},
  \bibinfo {author} {\bibfnamefont {Y.}~\bibnamefont {Tao}}, \ and\ \bibinfo
  {author} {\bibfnamefont {X.}~\bibnamefont {Fu}},\ }\href {\doibase
  10.1007/s11538-018-0445-z} {\bibfield  {journal} {\bibinfo  {journal} {Bull.
  Math. Biol.}\ }\textbf {\bibinfo {volume} {80}},\ \bibinfo {pages} {2049}
  (\bibinfo {year} {2018})}\BibitemShut {NoStop}%
\bibitem [{\citenamefont {Mieghem}\ and\ \citenamefont
  {Liu}(2019)}]{Mieghem2019}%
  \BibitemOpen
  \bibfield  {author} {\bibinfo {author} {\bibfnamefont {P.~V.}\ \bibnamefont
  {Mieghem}}\ and\ \bibinfo {author} {\bibfnamefont {Q.}~\bibnamefont {Liu}},\
  }\href {\doibase 10.1103/physreve.100.022317} {\bibfield  {journal} {\bibinfo
   {journal} {Phys. Rev. E}\ }\textbf {\bibinfo {volume} {100}},\ \bibinfo
  {pages} {022317} (\bibinfo {year} {2019})}\BibitemShut {NoStop}%
\bibitem [{\citenamefont {de~Arruda}\ \emph {et~al.}(2020)\citenamefont
  {de~Arruda}, \citenamefont {Petri}, \citenamefont {Rodrigues},\ and\
  \citenamefont {Moreno}}]{Arruda2020}%
  \BibitemOpen
  \bibfield  {author} {\bibinfo {author} {\bibfnamefont {G.~F.}\ \bibnamefont
  {de~Arruda}}, \bibinfo {author} {\bibfnamefont {G.}~\bibnamefont {Petri}},
  \bibinfo {author} {\bibfnamefont {F.~A.}\ \bibnamefont {Rodrigues}}, \ and\
  \bibinfo {author} {\bibfnamefont {Y.}~\bibnamefont {Moreno}},\ }\href
  {\doibase 10.1103/physrevresearch.2.013046} {\bibfield  {journal} {\bibinfo
  {journal} {Phys. Rev. Research}\ }\textbf {\bibinfo {volume} {2}},\ \bibinfo
  {pages} {013046} (\bibinfo {year} {2020})}\BibitemShut {NoStop}%
\bibitem [{\citenamefont {Starnini}\ \emph {et~al.}(2017)\citenamefont
  {Starnini}, \citenamefont {Gleeson},\ and\ \citenamefont
  {Bogu{\~{n}}{\'{a}}}}]{Starnini2017}%
  \BibitemOpen
  \bibfield  {author} {\bibinfo {author} {\bibfnamefont {M.}~\bibnamefont
  {Starnini}}, \bibinfo {author} {\bibfnamefont {J.~P.}\ \bibnamefont
  {Gleeson}}, \ and\ \bibinfo {author} {\bibfnamefont {M.}~\bibnamefont
  {Bogu{\~{n}}{\'{a}}}},\ }\href {\doibase 10.1103/physrevlett.118.128301}
  {\bibfield  {journal} {\bibinfo  {journal} {Phys. Rev. Lett.}\ }\textbf
  {\bibinfo {volume} {118}},\ \bibinfo {pages} {128301} (\bibinfo {year}
  {2017})}\BibitemShut {NoStop}%
\bibitem [{\citenamefont {Rausand}\ and\ \citenamefont
  {H{\o}yland}(2004)}]{SystemReliabilityTheory}%
  \BibitemOpen
  \bibfield  {author} {\bibinfo {author} {\bibfnamefont {M.}~\bibnamefont
  {Rausand}}\ and\ \bibinfo {author} {\bibfnamefont {A.}~\bibnamefont
  {H{\o}yland}},\ }\href@noop {} {\emph {\bibinfo {title} {System Reliability
  Theory: Models, Statistical Methods, and Applications}}}\ (\bibinfo
  {publisher} {John Wiley \& Sons, New York},\ \bibinfo {year}
  {2004})\BibitemShut {NoStop}%
\bibitem [{200(2006)}]{2006a}%
  \BibitemOpen
  \href {\doibase 10.1007/0-387-34232-x} {\emph {\bibinfo {title} {Stochastic
  Ageing and Dependence for Reliability}}}\ (\bibinfo  {publisher} {Springer
  New York},\ \bibinfo {year} {2006})\BibitemShut {NoStop}%
\bibitem [{\citenamefont {Rocchi}(2017)}]{Rocchi2017}%
  \BibitemOpen
  \bibfield  {author} {\bibinfo {author} {\bibfnamefont {P.}~\bibnamefont
  {Rocchi}},\ }\href {\doibase 10.1007/978-3-319-57472-1} {\emph {\bibinfo
  {title} {Reliability is a New Science}}}\ (\bibinfo  {publisher} {Springer
  International Publishing},\ \bibinfo {year} {2017})\BibitemShut {NoStop}%
\bibitem [{\citenamefont {Gfeller}\ and\ \citenamefont
  {Rios}(2007)}]{Gfeller2007}%
  \BibitemOpen
  \bibfield  {author} {\bibinfo {author} {\bibfnamefont {D.}~\bibnamefont
  {Gfeller}}\ and\ \bibinfo {author} {\bibfnamefont {P.~D.~L.}\ \bibnamefont
  {Rios}},\ }\href {\doibase 10.1103/physrevlett.99.038701} {\bibfield
  {journal} {\bibinfo  {journal} {Phys. Rev. Lett.}\ }\textbf {\bibinfo
  {volume} {99}},\ \bibinfo {pages} {038701} (\bibinfo {year}
  {2007})}\BibitemShut {NoStop}%
\bibitem [{\citenamefont {Gfeller}\ and\ \citenamefont
  {Rios}(2008)}]{Gfeller2008}%
  \BibitemOpen
  \bibfield  {author} {\bibinfo {author} {\bibfnamefont {D.}~\bibnamefont
  {Gfeller}}\ and\ \bibinfo {author} {\bibfnamefont {P.~D.~L.}\ \bibnamefont
  {Rios}},\ }\href {\doibase 10.1103/physrevlett.100.174104} {\bibfield
  {journal} {\bibinfo  {journal} {Phys. Rev. Lett.}\ }\textbf {\bibinfo
  {volume} {100}},\ \bibinfo {pages} {174104} (\bibinfo {year}
  {2008})}\BibitemShut {NoStop}%
\bibitem [{\citenamefont {Chen}\ \emph {et~al.}(2010)\citenamefont {Chen},
  \citenamefont {Hou}, \citenamefont {Xin},\ and\ \citenamefont
  {Yan}}]{Chen2010}%
  \BibitemOpen
  \bibfield  {author} {\bibinfo {author} {\bibfnamefont {H.}~\bibnamefont
  {Chen}}, \bibinfo {author} {\bibfnamefont {Z.}~\bibnamefont {Hou}}, \bibinfo
  {author} {\bibfnamefont {H.}~\bibnamefont {Xin}}, \ and\ \bibinfo {author}
  {\bibfnamefont {Y.}~\bibnamefont {Yan}},\ }\href {\doibase
  10.1103/physreve.82.011107} {\bibfield  {journal} {\bibinfo  {journal} {Phys.
  Rev. E}\ }\textbf {\bibinfo {volume} {82}},\ \bibinfo {pages} {011107}
  (\bibinfo {year} {2010})}\BibitemShut {NoStop}%
\bibitem [{\citenamefont {Shen}\ \emph {et~al.}(2011)\citenamefont {Shen},
  \citenamefont {Chen}, \citenamefont {Hou},\ and\ \citenamefont
  {Xin}}]{Shen2011}%
  \BibitemOpen
  \bibfield  {author} {\bibinfo {author} {\bibfnamefont {C.}~\bibnamefont
  {Shen}}, \bibinfo {author} {\bibfnamefont {H.}~\bibnamefont {Chen}}, \bibinfo
  {author} {\bibfnamefont {Z.}~\bibnamefont {Hou}}, \ and\ \bibinfo {author}
  {\bibfnamefont {H.}~\bibnamefont {Xin}},\ }\href {\doibase
  10.1103/physreve.83.066109} {\bibfield  {journal} {\bibinfo  {journal} {Phys.
  Rev. E}\ }\textbf {\bibinfo {volume} {83}},\ \bibinfo {pages} {066109}
  (\bibinfo {year} {2011})}\BibitemShut {NoStop}%
\bibitem [{\citenamefont {W.O.~Kermack}(1927)}]{1927a}%
  \BibitemOpen
  \bibfield  {author} {\bibinfo {author} {\bibfnamefont {A.~M.}\ \bibnamefont
  {W.O.~Kermack}},\ }\href {\doibase 10.1098/rspa.1927.0118} {\bibfield
  {journal} {\bibinfo  {journal} {Proc. Royal Soc. London}\ }\textbf {\bibinfo
  {volume} {115}},\ \bibinfo {pages} {700} (\bibinfo {year}
  {1927})}\BibitemShut {NoStop}%
\bibitem [{\citenamefont {Brauer}\ \emph {et~al.}(2019)\citenamefont {Brauer},
  \citenamefont {Castillo-Chavez},\ and\ \citenamefont {Feng}}]{Brauer2019}%
  \BibitemOpen
  \bibfield  {author} {\bibinfo {author} {\bibfnamefont {F.}~\bibnamefont
  {Brauer}}, \bibinfo {author} {\bibfnamefont {C.}~\bibnamefont
  {Castillo-Chavez}}, \ and\ \bibinfo {author} {\bibfnamefont {Z.}~\bibnamefont
  {Feng}},\ }\href
  {https://www.ebook.de/de/product/37134897/fred_brauer_carlos_castillo_chavez_zhilan_feng_mathematical_models_in_epidemiology.html}
  {\emph {\bibinfo {title} {Mathematical Models in Epidemiology}}}\ (\bibinfo
  {publisher} {Springer New York},\ \bibinfo {year} {2019})\BibitemShut
  {NoStop}%
\bibitem [{\citenamefont {Du}\ and\ \citenamefont
  {Sun}(2020)}]{Du_2020reliability}%
  \BibitemOpen
  \bibfield  {author} {\bibinfo {author} {\bibfnamefont {Y.-M.}\ \bibnamefont
  {Du}}\ and\ \bibinfo {author} {\bibfnamefont {C.-P.}\ \bibnamefont {Sun}},\
  }\href {\doibase 10.1360/tb-2020-0366} {\bibfield  {journal} {\bibinfo
  {journal} {Chin. Sci. Bull.}\ }\textbf {\bibinfo {volume} {65}},\ \bibinfo
  {pages} {2356} (\bibinfo {year} {2020})}\BibitemShut {NoStop}%
\bibitem [{\citenamefont {Du}\ \emph {et~al.}(2020)\citenamefont {Du},
  \citenamefont {Ma}, \citenamefont {Wei}, \citenamefont {Guan},\ and\
  \citenamefont {Sun}}]{Du2020}%
  \BibitemOpen
  \bibfield  {author} {\bibinfo {author} {\bibfnamefont {Y.-M.}\ \bibnamefont
  {Du}}, \bibinfo {author} {\bibfnamefont {Y.-H.}\ \bibnamefont {Ma}}, \bibinfo
  {author} {\bibfnamefont {Y.-F.}\ \bibnamefont {Wei}}, \bibinfo {author}
  {\bibfnamefont {X.}~\bibnamefont {Guan}}, \ and\ \bibinfo {author}
  {\bibfnamefont {C.~P.}\ \bibnamefont {Sun}},\ }\href {\doibase
  10.1103/physreve.101.012106} {\bibfield  {journal} {\bibinfo  {journal}
  {Phys. Rev. E}\ }\textbf {\bibinfo {volume} {101}},\ \bibinfo {pages}
  {012106} (\bibinfo {year} {2020})}\BibitemShut {NoStop}%
\bibitem [{sup()}]{supplementarymaterials}%
  \BibitemOpen
  \href@noop {} {\enquote {\bibinfo {title} {Supplementary materials},}\
  }\BibitemShut {NoStop}%
\bibitem [{\citenamefont {Bailey}(1975)}]{Bailey1975}%
  \BibitemOpen
  \bibfield  {author} {\bibinfo {author} {\bibfnamefont {N.~T.~J.}\
  \bibnamefont {Bailey}},\ }\href@noop {} {\emph {\bibinfo {title} {The
  Mathematical Theory of Infectious Diseases and Its Applications}}}\ (\bibinfo
   {publisher} {Griffin, London},\ \bibinfo {year} {1975})\BibitemShut
  {NoStop}%
\bibitem [{sel()}]{selfconsistentequation}%
  \BibitemOpen
  \href@noop {} {\enquote {\bibinfo {title} {In supplementary materials, we
  give the self-consistent equation with the infection rate
  $\beta(\tau_{0},\tau_{1})$ dependent on the susceptible duration
  $\tau_{0}$.}}\ }\BibitemShut {NoStop}%
\bibitem [{sin()}]{sincetheproduct}%
  \BibitemOpen
  \href@noop {} {\enquote {\bibinfo {title} {The product
  {$\Theta\bar{\tau}_{1}$} does not rely on $\bar{\tau}_{1}$, and is determined
  by {$\Upsilon$} and {$P(k)$} according to the self-consistent equation},}\
  }\BibitemShut {NoStop}%
\bibitem [{\citenamefont {Mieghem}(2006)}]{Mieghem2006}%
  \BibitemOpen
  \bibfield  {author} {\bibinfo {author} {\bibfnamefont {P.~V.}\ \bibnamefont
  {Mieghem}},\ }\href {\doibase 10.1017/cbo9780511616488} {\emph {\bibinfo
  {title} {Performance Analysis of Communications Networks and Systems}}}\
  (\bibinfo  {publisher} {Cambridge University Press},\ \bibinfo {year}
  {2006})\BibitemShut {NoStop}%
\bibitem [{\citenamefont {Catanzaro}\ \emph {et~al.}(2005)\citenamefont
  {Catanzaro}, \citenamefont {Bogu{\~{n}}{\'{a}}},\ and\ \citenamefont
  {Pastor-Satorras}}]{Catanzaro2005}%
  \BibitemOpen
  \bibfield  {author} {\bibinfo {author} {\bibfnamefont {M.}~\bibnamefont
  {Catanzaro}}, \bibinfo {author} {\bibfnamefont {M.}~\bibnamefont
  {Bogu{\~{n}}{\'{a}}}}, \ and\ \bibinfo {author} {\bibfnamefont
  {R.}~\bibnamefont {Pastor-Satorras}},\ }\href {\doibase
  10.1103/physreve.71.027103} {\bibfield  {journal} {\bibinfo  {journal} {Phys.
  Rev. E}\ }\textbf {\bibinfo {volume} {71}},\ \bibinfo {pages} {027103}
  (\bibinfo {year} {2005})}\BibitemShut {NoStop}%
\bibitem [{\citenamefont {Castellano}\ and\ \citenamefont
  {Pastor-Satorras}(2020)}]{Castellano2020}%
  \BibitemOpen
  \bibfield  {author} {\bibinfo {author} {\bibfnamefont {C.}~\bibnamefont
  {Castellano}}\ and\ \bibinfo {author} {\bibfnamefont {R.}~\bibnamefont
  {Pastor-Satorras}},\ }\href {\doibase 10.1103/physrevx.10.011070} {\bibfield
  {journal} {\bibinfo  {journal} {Phys. Rev. X}\ }\textbf {\bibinfo {volume}
  {10}},\ \bibinfo {pages} {011070} (\bibinfo {year} {2020})}\BibitemShut
  {NoStop}%
\end{thebibliography}%


\begin{thebibliography}{7}%
\makeatletter
\providecommand \@ifxundefined [1]{%
 \@ifx{#1\undefined}
}%
\providecommand \@ifnum [1]{%
 \ifnum #1\expandafter \@firstoftwo
 \else \expandafter \@secondoftwo
 \fi
}%
\providecommand \@ifx [1]{%
 \ifx #1\expandafter \@firstoftwo
 \else \expandafter \@secondoftwo
 \fi
}%
\providecommand \natexlab [1]{#1}%
\providecommand \enquote  [1]{``#1''}%
\providecommand \bibnamefont  [1]{#1}%
\providecommand \bibfnamefont [1]{#1}%
\providecommand \citenamefont [1]{#1}%
\providecommand \href@noop [0]{\@secondoftwo}%
\providecommand \href [0]{\begingroup \@sanitize@url \@href}%
\providecommand \@href[1]{\@@startlink{#1}\@@href}%
\providecommand \@@href[1]{\endgroup#1\@@endlink}%
\providecommand \@sanitize@url [0]{\catcode `\\12\catcode `\$12\catcode
  `\&12\catcode `\#12\catcode `\^12\catcode `\_12\catcode `\%12\relax}%
\providecommand \@@startlink[1]{}%
\providecommand \@@endlink[0]{}%
\providecommand \url  [0]{\begingroup\@sanitize@url \@url }%
\providecommand \@url [1]{\endgroup\@href {#1}{\urlprefix }}%
\providecommand \urlprefix  [0]{URL }%
\providecommand \Eprint [0]{\href }%
\providecommand \doibase [0]{http://dx.doi.org/}%
\providecommand \selectlanguage [0]{\@gobble}%
\providecommand \bibinfo  [0]{\@secondoftwo}%
\providecommand \bibfield  [0]{\@secondoftwo}%
\providecommand \translation [1]{[#1]}%
\providecommand \BibitemOpen [0]{}%
\providecommand \bibitemStop [0]{}%
\providecommand \bibitemNoStop [0]{.\EOS\space}%
\providecommand \EOS [0]{\spacefactor3000\relax}%
\providecommand \BibitemShut  [1]{\csname bibitem#1\endcsname}%
\let\auto@bib@innerbib\@empty
\bibitem [{\citenamefont {Brauer}\ \emph {et~al.}(2019)\citenamefont {Brauer},
  \citenamefont {Castillo-Chavez},\ and\ \citenamefont {Feng}}]{Brauer2019}%
  \BibitemOpen
  \bibfield  {author} {\bibinfo {author} {\bibfnamefont {F.}~\bibnamefont
  {Brauer}}, \bibinfo {author} {\bibfnamefont {C.}~\bibnamefont
  {Castillo-Chavez}}, \ and\ \bibinfo {author} {\bibfnamefont {Z.}~\bibnamefont
  {Feng}},\ }\href
  {https://www.ebook.de/de/product/37134897/fred_brauer_carlos_castillo_chavez_zhilan_feng_mathematical_models_in_epidemiology.html}
  {\emph {\bibinfo {title} {Mathematical Models in Epidemiology}}}\ (\bibinfo
  {publisher} {Springer New York},\ \bibinfo {year} {2019})\BibitemShut
  {NoStop}%
\bibitem [{\citenamefont {Du}\ \emph {et~al.}(2020)\citenamefont {Du},
  \citenamefont {Ma}, \citenamefont {Wei}, \citenamefont {Guan},\ and\
  \citenamefont {Sun}}]{Du2020}%
  \BibitemOpen
  \bibfield  {author} {\bibinfo {author} {\bibfnamefont {Y.-M.}\ \bibnamefont
  {Du}}, \bibinfo {author} {\bibfnamefont {Y.-H.}\ \bibnamefont {Ma}}, \bibinfo
  {author} {\bibfnamefont {Y.-F.}\ \bibnamefont {Wei}}, \bibinfo {author}
  {\bibfnamefont {X.}~\bibnamefont {Guan}}, \ and\ \bibinfo {author}
  {\bibfnamefont {C.~P.}\ \bibnamefont {Sun}},\ }\href {\doibase
  10.1103/physreve.101.012106} {\bibfield  {journal} {\bibinfo  {journal}
  {Phys. Rev. E}\ }\textbf {\bibinfo {volume} {101}},\ \bibinfo {pages}
  {012106} (\bibinfo {year} {2020})}\BibitemShut {NoStop}%
\bibitem [{\citenamefont {Du}\ and\ \citenamefont
  {Sun}(2020)}]{Du_2020reliability}%
  \BibitemOpen
  \bibfield  {author} {\bibinfo {author} {\bibfnamefont {Y.-M.}\ \bibnamefont
  {Du}}\ and\ \bibinfo {author} {\bibfnamefont {C.-P.}\ \bibnamefont {Sun}},\
  }\href {\doibase 10.1360/tb-2020-0366} {\bibfield  {journal} {\bibinfo
  {journal} {Chin. Sci. Bull.}\ }\textbf {\bibinfo {volume} {65}},\ \bibinfo
  {pages} {2356} (\bibinfo {year} {2020})}\BibitemShut {NoStop}%
\bibitem [{\citenamefont {Barrat}\ \emph {et~al.}(2012)\citenamefont {Barrat},
  \citenamefont {Barthelemy},\ and\ \citenamefont
  {Vespignani}}]{MarcBarthelemy2012}%
  \BibitemOpen
  \bibfield  {author} {\bibinfo {author} {\bibfnamefont {A.}~\bibnamefont
  {Barrat}}, \bibinfo {author} {\bibfnamefont {M.}~\bibnamefont {Barthelemy}},
  \ and\ \bibinfo {author} {\bibfnamefont {A.}~\bibnamefont {Vespignani}},\
  }\href
  {https://www.ebook.de/de/product/7871347/marc_barthelemy_alessandro_vespignani_dynamical_processes_on_complex_networks.html}
  {\emph {\bibinfo {title} {Dynamical Processes on Complex Networks}}}\
  (\bibinfo  {publisher} {Cambridge University Press},\ \bibinfo {year}
  {2012})\BibitemShut {NoStop}%
\bibitem [{\citenamefont {Pastor-Satorras}\ and\ \citenamefont
  {Vespignani}(2001)}]{Pastor-Satorras2001}%
  \BibitemOpen
  \bibfield  {author} {\bibinfo {author} {\bibfnamefont {R.}~\bibnamefont
  {Pastor-Satorras}}\ and\ \bibinfo {author} {\bibfnamefont {A.}~\bibnamefont
  {Vespignani}},\ }\href {\doibase 10.1103/physrevlett.86.3200} {\bibfield
  {journal} {\bibinfo  {journal} {Phys. Rev. Lett.}\ }\textbf {\bibinfo
  {volume} {86}},\ \bibinfo {pages} {3200} (\bibinfo {year}
  {2001})}\BibitemShut {NoStop}%
\bibitem [{\citenamefont {Li}\ \emph {et~al.}(2012)\citenamefont {Li},
  \citenamefont {van~de Bovenkamp},\ and\ \citenamefont {Mieghem}}]{Li2012}%
  \BibitemOpen
  \bibfield  {author} {\bibinfo {author} {\bibfnamefont {C.}~\bibnamefont
  {Li}}, \bibinfo {author} {\bibfnamefont {R.}~\bibnamefont {van~de
  Bovenkamp}}, \ and\ \bibinfo {author} {\bibfnamefont {P.~V.}\ \bibnamefont
  {Mieghem}},\ }\href {\doibase 10.1103/physreve.86.026116} {\bibfield
  {journal} {\bibinfo  {journal} {Phys. Rev. E}\ }\textbf {\bibinfo {volume}
  {86}},\ \bibinfo {pages} {026116} (\bibinfo {year} {2012})}\BibitemShut
  {NoStop}%
\bibitem [{\citenamefont {Catanzaro}\ \emph {et~al.}(2005)\citenamefont
  {Catanzaro}, \citenamefont {Bogu{\~{n}}{\'{a}}},\ and\ \citenamefont
  {Pastor-Satorras}}]{Catanzaro2005}%
  \BibitemOpen
  \bibfield  {author} {\bibinfo {author} {\bibfnamefont {M.}~\bibnamefont
  {Catanzaro}}, \bibinfo {author} {\bibfnamefont {M.}~\bibnamefont
  {Bogu{\~{n}}{\'{a}}}}, \ and\ \bibinfo {author} {\bibfnamefont
  {R.}~\bibnamefont {Pastor-Satorras}},\ }\href {\doibase
  10.1103/physreve.71.027103} {\bibfield  {journal} {\bibinfo  {journal} {Phys.
  Rev. E}\ }\textbf {\bibinfo {volume} {71}},\ \bibinfo {pages} {027103}
  (\bibinfo {year} {2005})}\BibitemShut {NoStop}%
\end{thebibliography}%

\end{document}


\title{Supplementary Materials: Hierarchical Coarse-grained Approach to the
Duration-dependent Spreading Dynamics on Complex Networks}
\author{Jin-Fu Chen}
\address{Beijing Computational Science Research Center, Beijing 100193, China}
\address{Graduate School of China Academy of Engineering Physics, Beijing,
100193, China}
\author{Yi-Mu Du}
\address{Graduate School of China Academy of Engineering Physics, Beijing,
100193, China}
\author{Hui Dong}
\email{hdong@gscaep.ac.cn}

\address{Graduate School of China Academy of Engineering Physics, Beijing,
100193, China}
\author{Chang-Pu Sun}
\email{cpsun@csrc.ac.cn}

\address{Beijing Computational Science Research Center, Beijing 100193, China}
\address{Graduate School of China Academy of Engineering Physics, Beijing,
100193, China}
\date{\today}

\maketitle
The document is devoted to providing detailed discussions and derivations
to support the main content. In Sec. \ref{sec:the-spreading-dynamics},
we build the microscopic spreading model by introducing the probability
density function (PDF) of the duration for each node in the network.
In Sec. \ref{sec:Derive-the-mesoscopic}, we show the emergence of
the duration coarse-grained (DCG) approach to obtain the mesoscopic
and macroscopic models. In Sec. \ref{sec:Duration-dependent-coarse-graine},
the DCG approach is applied to the susceptible-infected-susceptible
(SIS) model. In Sec. \ref{sec:Relation-to-the}, we show the macroscopic
SIS model recovers to the standard \citep{Brauer2019} and the extended
compartmental models \citep{Du2020}. In Sec. \ref{sec:The-steady-state},
we solve the steady state of the mesoscopic SIS model in an uncorrelated
network with duration-dependent recovery and infection rates. In Sec.
\ref{sec:Monte-Carlo-Simulation}, we provide the details of the continuous-time
Monte Carlo simulation of the SIS model in an uncorrelated scale-free
network.

\section{spreading dynamics in microscopic model\label{sec:the-spreading-dynamics}}

In the basic processes, the node transforms from one state to another
state. In the microscopic model, the probability distribution $P_{l,i}(t)$
describes the probability of the node $l$ staying in the state $i$,
and satisfies the normalization condition $\sum_{i}P_{l,i}(t)=1$.
We assume the states of different nodes are uncorrelated: the probability
of the node $l$ in the state $i$ and the node $m$ in the state
$j$ can be written in the product form $P_{l,i}(t)\times P_{m,j}(t)$.

\subsection{Probability density function $\rho_{l,i}(\tau_{i},t)$ of the duration
and its equation for the evolution}

Under the uncorrelated assumption, we introduce the probability density
function (PDF) $\rho_{l,i}(\tau_{i},t)$ for each node with the duration
$\tau_{i}$ on the state $i$ to describe the state of the network.
For the node $l$, the probability in the state $i$ with the duration
between $\tau_{i}$ and $\tau_{i}+\delta\tau_{i}$ is $\rho_{l,i}(\tau_{i},t)\delta\tau_{i}$.
The probability $P_{l,i}(t)$ is equal to the integral

\begin{equation}
P_{l,i}(t)=\int_{0}^{\infty}\rho_{l,i}(\tau_{i},t)d\tau_{i}.
\end{equation}

The total transformation rate $\Gamma_{l,i}(\tau_{i},t)$ from the
state $i$ to the other states is

\begin{equation}
\Gamma_{l,i}(\tau_{i},t)=\sum_{i^{\prime}}\gamma_{l,i^{\prime}i}(\tau_{i},t),\label{eq:totaltransformationrate}
\end{equation}
where $\gamma_{l,i^{\prime}i}(\tau_{i},t)$ is the transformation
rate from the state $i$ to the state $i^{\prime}$. In a small time
step $dt$, the node $l$ transforms from $i$ to other states with
the conditional probability $\Gamma_{l,i}(\tau_{i},t)dt$. At the
time $t+dt$, the probability in the state $i$ with the duration
between $\tau_{i}+dt$ and $\tau_{i}+\delta\tau_{i}+dt$ is $\rho_{l,i}(\tau_{i}+dt,t+dt)\delta\tau_{i}$.
The change of the probability is caused by the transformation, namely,
\begin{equation}
\rho_{l,i}(\tau_{i}+dt,t+dt)\delta\tau_{i}-\rho_{l,i}(\tau_{i},t)\delta\tau_{i}=-\Gamma_{l,i}(\tau_{i},t)dt\rho_{l,i}(\tau_{i},t)\delta\tau_{i}.
\end{equation}
With the above equation, we obtain the differential equation for the
PDF as

\begin{equation}
\frac{\partial\rho_{l,i}(\tau_{i},t)}{\partial\tau_{i}}+\frac{\partial\rho_{l,i}(\tau_{i},t)}{\partial t}=-\Gamma_{l,i}(\tau_{i},t)\rho_{l,i}(\tau_{i},t).\label{eq:PDFdifferentialeq}
\end{equation}

\subsection{The transformation rate $\gamma_{l,i^{\prime}i}(\tau_{i},t)$}

The transformation rate $\gamma_{l,i^{\prime}i}(\tau_{i},t)$ relates
to the basic processes with the transformation from the state $i$
to the state $i^{\prime}$. In an aging process $i\overset{\alpha_{i^{\prime},i}}{\longrightarrow}i^{\prime}$,
the contribution to the transformation rate is given directly by the
transition rate 
\begin{equation}
\gamma_{l,i^{\prime}i}^{(\mathrm{aging})}(\tau_{i},t)=\alpha_{i^{\prime},i}(\tau_{i}).\label{eq:aginggamma}
\end{equation}
In a contact process $i+j\overset{\beta_{i^{\prime}j^{\prime},ij}}{\longrightarrow}i^{\prime}+j^{\prime}$,
the transformation depends on the states and the duration of all the
neighbors $m$ as

\begin{equation}
\gamma_{l,i^{\prime}i}^{(\mathrm{contact})}(\tau_{i}^{(l)},t)=\sum_{m,j,j^{\prime}}A_{lm}\int_{0}^{\infty}\beta_{i^{\prime}j^{\prime},ij}(\tau_{i}^{(l)},\tau_{j}^{(m)})\rho_{m,j}(\tau_{j}^{(m)},t)d\tau_{j}^{(m)}.\label{eq:contactaging}
\end{equation}
where $A_{lm}$ is the adjacency matrix of the network: $A_{lm}=1$
if the nodes $l$ and $m$ are linked, otherwise $A_{lm}=0$.

Including the contribution from both the aging and the contact processes,
the overall transformation rate follows as

\begin{equation}
\gamma_{l,i^{\prime}i}(\tau_{i}^{(l)},t)=\alpha_{i^{\prime},i}(\tau_{i}^{(l)})+\sum_{m,j,j^{\prime}}A_{lm}\int_{0}^{\infty}\beta_{i^{\prime}j^{\prime},ij}(\tau_{i}^{(l)},\tau_{j}^{(m)})\rho_{m,j}(\tau_{j}^{(m)},t)d\tau_{j}^{(m)},
\end{equation}
which is Eq. (3) in the main content.

\subsection{Connecting condition}

For the node $l$, we define the flux from the state $i^{\prime}$
to the state $i$ as

\begin{equation}
\phi_{l,ii^{\prime}}(t)=\int_{0}^{\infty}\gamma_{l,ii^{\prime}}(\tau_{i^{\prime}},t)\times\rho_{l,i^{\prime}}(\tau_{i^{\prime}},t)d\tau_{i^{\prime}},
\end{equation}
which is the probability for the transformation from the state $i^{\prime}$
to the state $i$ in unit time. The total flux to the state $i$ from
all other states is

\begin{equation}
\Phi_{l,i}(t)=\sum_{i^{\prime}}\phi_{l,ii^{\prime}}(t).
\end{equation}
In the small time step $dt$, the probability $\rho_{l,i}(0,t)dt$
of the transformation to the state $i$ is equal to $\Phi_{l,i}(t)dt$
due to the conservation of the probability as

\begin{equation}
\rho_{l,i}(0,t)=\Phi_{l,i}(t).
\end{equation}

The change of the probability relates to the fluxes as

\begin{align}
\frac{dP_{l,i}(t)}{dt} & =\int_{0}^{\infty}\frac{\partial\rho_{l,i}(\tau_{i},t)}{\partial t}d\tau_{i}\nonumber \\
 & =\int_{0}^{\infty}[-\frac{\partial\rho_{l,i}(\tau_{i},t)}{\partial\tau_{i}}-\Gamma_{l,i}(\tau_{i},t)\rho_{l,i}(\tau_{i},t)]d\tau_{i}\nonumber \\
 & =\rho_{l,i}(0,t)-\rho_{l,i}(\infty,t)-\int_{0}^{\infty}\Gamma_{l,i}(\tau_{i},t)\rho_{l,i}(\tau_{i},t)d\tau_{i}\nonumber \\
 & =\sum_{i^{\prime}}[\phi_{l,ii^{\prime}}(t)-\phi_{l,i^{\prime}i}(t)].
\end{align}
In the above derivation, we have used the equation of the evolution
by Eq. (\ref{eq:PDFdifferentialeq}) and the condition $\rho_{l,i}(\infty,t)=0$.
For constant transition rates with $\alpha_{i^{\prime},i}(\tau_{i})=\alpha_{i^{\prime},i}$
and $\beta_{i^{\prime}j^{\prime},ij}(\tau_{i},\tau_{j})=\beta_{i^{\prime}j^{\prime},ij}$,
the flux $\phi_{l,i^{\prime}i}(t)$ is directly given by the probability
as

\begin{equation}
\phi_{l,i^{\prime}i}(t)=[\alpha_{i^{\prime},i}+\sum_{m,j^{\prime},j}A_{lm}\beta_{i^{\prime}j^{\prime},ij}P_{m,j}(t)]P_{l,i}(t).
\end{equation}

\section{Hierarchical duration coarse-grained approach\label{sec:Derive-the-mesoscopic}}

Typically, the spreading dynamics can be evaluated through the populations
of different states. The duration coarse-grained (DCG) approach enables
us to derive the coarse-grained models with the populations from the
microscopic model with the probabilities. Here, we supplement the
derivation of the duration coarse-grained approach in the main content
and show the hierarchy among the microscopic, mesoscopic and macroscopic
models.

\subsection{Mesoscopic model}

In the mesoscopic model, the state of the network is described by
the coarse-grained PDF $\rho_{k,i}(\tau_{i},t)$ for the $k$-degree
nodes, which relates to the PDF of each node as

\begin{equation}
\rho_{k,i}(\tau_{i},t)=\sum_{l}\delta_{k,k_{l}}\rho_{l,i}(\tau_{i},t)/n_{k},\label{eq:DDFfki}
\end{equation}
where $k_{l}$ is the degree of the node $l$ and $n_{k}$ is the
population of the $k$-degree nodes. The differential equation of
the coarse-grained PDF is

\begin{equation}
\frac{\partial\rho_{k,i}(\tau_{i},t)}{\partial\tau_{i}}+\frac{\partial\rho_{k,i}(\tau_{i},t)}{\partial t}=-\sum_{l}\frac{\delta_{k,k_{l}}\Gamma_{l,i}(\tau_{i},t)}{n_{k}}\rho_{l,i}(\tau_{i},t).\label{eq:DDFdifferentialequation}
\end{equation}

The coarse-grained PDF is assumed identical for the nodes with the
same degree

\begin{equation}
\rho_{l,i}(\tau_{i},t)=\rho_{k_{l},i}(\tau_{i},t).\label{eq:rhol_i}
\end{equation}
By Eq. (\ref{eq:DDFdifferentialequation}), the corresponding transformation
rate for the $k$-degree nodes is

\begin{equation}
\Gamma_{k,i}(\tau_{i},t)=\sum_{l}\frac{\delta_{k,k_{l}}\Gamma_{l,i}(\tau_{i},t)}{n_{k}},
\end{equation}
and

\begin{equation}
\gamma_{k,i^{\prime}i}(\tau_{i},t)=\sum_{l}\frac{\delta_{k,k_{l}}\gamma_{l,i^{\prime}i}(\tau_{i},t)}{n_{k}}.\label{eq:gamma_kiprimei}
\end{equation}
Then, the differential equation of the coarse-grained PDF is rewritten
as

\begin{equation}
\frac{\partial\rho_{k,i}(\tau_{i},t)}{\partial\tau_{i}}+\frac{\partial\rho_{k,i}(\tau_{i},t)}{\partial t}=-\Gamma_{k,i}(\tau_{i},t)\rho_{k,i}(\tau_{i},t).\label{eq:differentialequationofDDFmesoscopic}
\end{equation}

Plugging the transformation rate $\gamma_{l,i^{\prime}i}(\tau_{i},t)$
of the node $l$ into Eq. (\ref{eq:gamma_kiprimei}), we obtain the
transformation rate $\gamma_{k,i^{\prime}i}(\tau_{i},t)$ for the
$k$-degree nodes in the main content as

\begin{align}
\gamma_{k,i^{\prime}i}(\tau_{i},t) & =\sum_{l}\frac{\delta_{k,k_{l}}}{n_{k}}[\alpha_{i^{\prime},i}(\tau_{i})+\sum_{m,j,j^{\prime}}A_{lm}\int_{0}^{\infty}\beta_{i^{\prime}j^{\prime},ij}(\tau_{i},\tau_{j})\rho_{m,j}(\tau_{j},t)d\tau_{j},].\nonumber \\
 & =\alpha_{i^{\prime},i}(\tau_{i})+\sum_{j,j^{\prime}}\sum_{k^{\prime}=1}^{\infty}\frac{\sum_{l,m}\left(\delta_{k,k_{l}}\delta_{k^{\prime},k_{m}}A_{lm}\right)}{n_{k}}\int_{0}^{\infty}\beta_{i^{\prime}j^{\prime},ij}(\tau_{i},\tau_{j})\rho_{k^{\prime},j}(\tau_{j},t)d\tau_{j}\label{eq:derivegammakiprimei}\\
 & =\alpha_{i^{\prime},i}(\tau_{i})+\sum_{j,j^{\prime}}\sum_{k^{\prime}=1}^{\infty}\frac{(1+\delta_{k,k^{\prime}})M_{kk^{\prime}}}{n_{k}}\int_{0}^{\infty}\beta_{i^{\prime}j^{\prime},ij}(\tau_{i},\tau_{j})\rho_{k^{\prime},j}(\tau_{j},t)d\tau_{j},
\end{align}
where $M_{kk^{\prime}}=\sum_{l,m}\left(\delta_{k,k_{l}}\delta_{k^{\prime},k_{m}}A_{lm}\right)/(1+\delta_{k,k^{\prime}})$
is the number of the edges linked two nodes with the degrees $k$
and $k^{\prime}$. We have used the identical assumption $\rho_{l,i}(\tau_{i},t)=\rho_{k_{l},i}(\tau_{i},t)$
in Eq. (\ref{eq:derivegammakiprimei}). For a $k$-degree node, the
conditional probability of having a $k^{\prime}$-degree neighbor
is described by the degree correlation $P(k^{\prime}|k)$, which is
explicitly determined by the edge number $M_{kk^{\prime}}$ as

\begin{equation}
P(k^{\prime}|k)=\frac{(1+\delta_{k,k^{\prime}})M_{kk^{\prime}}}{\sum_{k^{\prime}=1}^{\infty}(1+\delta_{k,k^{\prime}})M_{kk^{\prime}}}.
\end{equation}
The number of the edges linked to a $k$-degree node relates to the
number of $k$-degree nodes as 
\begin{equation}
\sum_{k^{\prime}=1}^{\infty}(1+\delta_{k,k^{\prime}})M_{kk^{\prime}}=kn_{k}.
\end{equation}
We obtain the transformation rate for the $k$-degree nodes as

\begin{equation}
\gamma_{k,i^{\prime}i}(\tau_{i},t)=\alpha_{i^{\prime},i}(\tau_{i})+k\sum_{j,j^{\prime}}\sum_{k^{\prime}=1}^{\infty}P(k^{\prime}|k)\int_{0}^{\infty}\beta_{i^{\prime}j^{\prime},ij}(\tau_{i},\tau_{j})\rho_{k^{\prime},j}(\tau_{j},t)d\tau_{j},\label{eq:gamma_kiprimeiresult}
\end{equation}
which is Eq. (4) in the main content.

According to Eq. (\ref{eq:DDFfki}), the connecting condition of the
coarse-grained PDF is $\rho_{k,i}(0,t)=\sum_{l}\delta_{k,k_{l}}\rho_{l}(0,t)/n_{k}$,
which leads to the mesoscopic flux as

\begin{equation}
\phi_{k,ii^{\prime}}(t)=\sum_{l}\delta_{k,k_{l}}\phi_{l,ii^{\prime}}(t).
\end{equation}
Under the identical assumption, the mesoscopic flux is determined
by the coarse-grained PDF as

\begin{align}
\phi_{k,ii^{\prime}}(t) & =\int_{0}^{\infty}\gamma_{k,ii^{\prime}}(\tau_{i^{\prime}},t)\rho_{k,i^{\prime}}(\tau_{i^{\prime}},t)d\tau_{i^{\prime}}.\label{eq:phikiiprime}
\end{align}
The total flux follows

\begin{equation}
\Phi_{k,i}(t)=\sum_{i^{\prime}}\phi_{k,ii^{\prime}}(t).
\end{equation}
The connecting condition of the coarse-grained PDF is rewritten as
\begin{equation}
\rho_{k,i}(0,t)=\Phi_{k,i}(t).\label{eq:connectingcondition}
\end{equation}

The change of the population of $k$-degree nodes in the state $i$
is obtained as

\begin{align}
\frac{dn_{k,i}(t)}{dt} & =n_{k}\sum_{i^{\prime}}[\phi_{k,ii^{\prime}}(t)-\phi_{k,i^{\prime}i}(t)].
\end{align}
For the constant transition rates $\alpha_{i^{\prime},i}(\tau_{i})=\alpha_{i^{\prime},i}$
and $\beta_{i^{\prime}j^{\prime},ij}(\tau_{i},\tau_{j})=\beta_{i^{\prime}j^{\prime},ij}$,
the mesoscopic flux is

\begin{equation}
\phi_{k,i^{\prime}i}(t)=[\alpha_{i^{\prime},i}+\sum_{k^{\prime}=1}^{\infty}kP(k^{\prime}|k)\sum_{j,j^{\prime}}\beta_{i^{\prime}j^{\prime},ij}P_{k^{\prime},j}(t)]P_{k,i}(t),
\end{equation}
where $P_{k,i}(t)=\int_{0}^{\infty}\rho_{k,i}(\tau_{i},t)d\tau_{i}$
is the probability of a $k$-degree node in the state $i$.

\subsection{Macroscopic model}

At the macroscopic level, we introduce the gross PDF to describe the
nodes in the state $i$ without distinguishing the degrees as

\begin{equation}
\rho_{i}(\tau_{i},t)=\sum_{k=1}^{\infty}P(k)\rho_{k,i}(\tau_{i},t).\label{eq:grossddfdefine}
\end{equation}
The differential equation of the gross PDF follows from Eq. (\ref{eq:differentialequationofDDFmesoscopic})
as

\begin{equation}
\frac{\partial\rho_{i}(\tau_{i},t)}{\partial\tau_{i}}+\frac{\partial\rho_{i}(\tau_{i},t)}{\partial t}=-\sum_{k=1}^{\infty}\Gamma_{k,i}(\tau_{i},t)P(k)\rho_{k,i}(\tau_{i},t).\label{eq:grossddf}
\end{equation}

For the homogeneous network with similar degrees of all nodes, the
PDF of each node approximates the same $\rho_{l,i}(\tau_{i},t)\simeq\rho_{i}(\tau_{i},t)$.
The right hand side of Eq. (\ref{eq:grossddf}) becomes $-\sum_{k=1}^{\infty}\Gamma_{k,i}(\tau_{i},t)P(k)\rho_{k,i}(\tau_{i},t)=-\sum_{k=1}^{\infty}[\Gamma_{k,i}(\tau_{i},t)P(k)]\rho_{i}(\tau_{i},t)$.
The corresponding transformation rate follows as 
\begin{equation}
\gamma_{i^{\prime}i}(\tau_{i},t)=\sum_{k}P(k)\gamma_{k,i^{\prime}i}(\tau_{i},t),\label{eq:gammamacroscopic}
\end{equation}
and

\begin{equation}
\Gamma_{i}(\tau_{i},t)=\sum_{i^{\prime}}\gamma_{i^{\prime}i}(\tau_{i},t).
\end{equation}
The differential equation of the gross PDF is rewritten as

\begin{equation}
\frac{\partial\rho_{i}(\tau_{i},t)}{\partial\tau_{i}}+\frac{\partial\rho_{i}(\tau_{i},t)}{\partial t}=-\Gamma_{i}(\tau_{i},t)\rho_{i}(\tau_{i},t).\label{eq:grossddfdifferentialeq}
\end{equation}

Plugging Eq. (\ref{eq:gamma_kiprimeiresult}) into $\gamma_{i^{\prime}i}(\tau_{i},t)=\sum_{k=1}^{\infty}P(k)\gamma_{k,i^{\prime}i}(\tau_{i},t)$,
we obtain the transformation rate $\gamma_{i^{\prime}i}(\tau_{i},t)$
from the state $i$ to the state $i^{\prime}$ as

\begin{equation}
\gamma_{i^{\prime}i}(\tau_{i},t)=\alpha_{i^{\prime},i}(\tau_{i})+\left\langle k\right\rangle \sum_{j,j^{\prime}}\int_{0}^{\infty}\beta_{i^{\prime}j^{\prime},ij}(\tau_{i},\tau_{j})\rho_{j}(\tau_{j},t)d\tau_{j},
\end{equation}
where we have used the normalization condition $\sum_{k^{\prime}}P(k^{\prime}|k)=1$.

The connecting condition of the gross PDF is
\begin{equation}
\rho_{i}(0,t)=\Phi_{i}(t),
\end{equation}
with $\Phi_{i}(t)=\sum_{i^{\prime}}\phi_{ii^{\prime}}(t)$ and $\phi_{ii^{\prime}}(t)=\int_{0}^{\infty}\gamma_{ii^{\prime}}(\tau_{i^{\prime}},t)\rho_{i^{\prime}}(\tau_{i^{\prime}},t)d\tau_{i^{\prime}}.$

The change of the population in the state $i$ is obtained as

\begin{align}
\frac{dN_{i}(t)}{dt} & =N_{T}\sum_{i^{\prime}}[\phi_{ii^{\prime}}(t)-\phi_{i^{\prime}i}(t)].
\end{align}
For the constant transition rates $\alpha_{i^{\prime},i}(\tau_{i})=\alpha_{i^{\prime},i}$
and $\beta_{i^{\prime}j^{\prime},ij}(\tau_{i},\tau_{j})=\beta_{i^{\prime}j^{\prime},ij}$,
the macroscopic flux is

\begin{equation}
\phi_{i^{\prime}i}(t)=[\alpha_{i^{\prime},i}+\left\langle k\right\rangle \sum_{j,j^{\prime}}\beta_{i^{\prime}j^{\prime},ij}P_{j}(t)]P_{i}(t),
\end{equation}
where $P_{i}(t)$ is the probability of a node in the state $i$.

\section{Duration coarse-grained approach to SIS model \label{sec:Duration-dependent-coarse-graine}}

In the SIS model, the nodes stay in the susceptible state $0$ or
the infected state $1$. The basic processes consist of an aging process
$1\overset{\alpha(\tau_{1})}{\longrightarrow}0$ with the recovery
rate $\alpha(\tau_{1})$ and a contact process $0+1\overset{\beta(\tau_{0},\tau_{1})}{\longrightarrow}1+1$
with the infection rate $\beta(\tau_{0},\tau_{1})$.

In the mesoscopic model, the node states in the network are described
by the coarse-grained PDF $\rho_{k,i}(\tau_{i},t),\:i=0,1$. That
of the $k$-degree nodes in the susceptible state satisfies

\begin{equation}
\frac{\partial\rho_{k,0}(\tau_{0},t)}{\partial\tau_{0}}+\frac{\partial\rho_{k,0}(\tau_{0},t)}{\partial t}=-\Gamma_{k,0}(\tau_{0},t)\rho_{k,0}(\tau_{0},t),\label{eq:DDFof SISmoel}
\end{equation}
where the transformation rate $\Gamma_{k,0}(\tau_{0},t)$ is induced
by the infection as

\begin{equation}
\Gamma_{k,0}(\tau_{0},t)=k\int_{0}^{\infty}\beta(\tau_{0},\tau_{1})\sum_{k^{\prime}=1}^{\infty}P(k^{\prime}|k)\rho_{k^{\prime},1}(\tau_{1},t)d\tau_{1}.
\end{equation}
The coarse-grained PDF of the $k$-degree nodes in the infected state
satisfies

\begin{equation}
\frac{\partial\rho_{k,1}(\tau_{1},t)}{\partial\tau_{1}}+\frac{\partial\rho_{k,1}(\tau_{1},t)}{\partial t}=-\alpha(\tau_{1})\rho_{k,1}(\tau_{1},t),\label{eq:DDFofSIS curing}
\end{equation}
where the transformation rate is the recovery rate $\alpha(\tau_{1})$.

The connecting condition is given by the flux $\rho_{k,i}(0,t)=\Phi_{k,i}(t)$,
where the fluxes are determined by the transformation rates as

\begin{equation}
\Phi_{k,1}(t)=\int_{0}^{\infty}\Gamma_{k,0}(\tau_{0},t)\rho_{k,0}(\tau_{0},t)d\tau_{0},
\end{equation}
and

\begin{equation}
\Phi_{k,0}(t)=\int_{0}^{\infty}\alpha(\tau_{1})\rho_{k,1}(\tau_{1},t)d\tau_{1}.\label{eq:Phik0}
\end{equation}

\section{Relation to the compartmental models\label{sec:Relation-to-the}}

In the following, we show the macroscopic model recovers to the compartmental
SIS model. The duration of all the susceptible and the infected individuals
is described by the gross PDFs as $\rho_{i}(\tau_{i},t)$ with $i=0,1$.
The network structure is coarse-grainedly described by the average
degree $\left\langle k\right\rangle =\sum_{k}kP(k)$.

The equations of the gross PDFs for the susceptible and the infected
states are obtained from Eq. (\ref{eq:grossddfdifferentialeq}) as
\begin{equation}
\frac{\partial\rho_{0}(\tau_{0},t)}{\partial\tau_{0}}+\frac{\partial\rho_{0}(\tau_{0},t)}{\partial t}=-\left\langle k\right\rangle [\int_{0}^{\infty}\beta(\tau_{0},\tau_{1})\rho_{1}(\tau_{1},t)d\tau_{1}]\rho_{0}(\tau_{0},t),
\end{equation}
 and 
\begin{equation}
\frac{\partial\rho_{1}(\tau_{1},t)}{\partial\tau_{1}}+\frac{\partial\rho_{1}(\tau_{1},t)}{\partial t}=-\alpha(\tau_{1})\rho_{1}(\tau_{1},t),\label{eq:f1differentialeq}
\end{equation}
The connecting conditions of DDFs are $\rho_{i}(0,t)=\Phi_{i}(t),\,i=0,1$,
with the fluxes determined as 
\begin{equation}
\Phi_{0}(t)=\int_{0}^{\infty}\alpha(\tau_{1})\rho_{1}(\tau_{1},t)d\tau_{1}\label{eq:Phi0}
\end{equation}
and 
\begin{equation}
\Phi_{1}(t)=\left\langle k\right\rangle \int_{0}^{\infty}\int_{0}^{\infty}\beta(\tau_{0},\tau_{1})\rho_{0}(\tau_{0},t)\rho_{1}(\tau_{1},t)d\tau_{0}d\tau_{1}.\label{eq:Phi1}
\end{equation}
The populations of the susceptible and the infected individuals are
$N_{i}(t)=N_{T}\int_{0}^{\infty}\rho_{i}(\tau_{i},t)d\tau_{i},\:i=0,1$,
which satisfy

\begin{align}
\frac{dN_{0}(t)}{dt} & =N_{T}[\Phi_{0}(t)-\Phi_{1}(t)]\label{eq:N0differential}\\
\frac{dN_{1}(t)}{dt} & =N_{T}[-\Phi_{0}(t)+\Phi_{1}(t)]\label{eq:N1differential}
\end{align}
The total population is $N_{T}=N_{0}(t)+N_{1}(t)$.

\subsection{The extended compartmental SIS model with integral-differential equations}

The extended SIS compartmental model requires the constant infection
rate $\beta(\tau_{0},\tau_{1})=\beta$, but the recovery rate $\alpha(\tau_{1})$
can be duration-dependent. The flux $\Phi_{1}(t)$ by Eq. (\ref{eq:Phi1})
is simplified as

\begin{equation}
\Phi_{1}(t)=\left\langle k\right\rangle \beta P_{0}(t)P_{1}(t).\label{eq:Phi1constantbeta}
\end{equation}

The formal solution of $\rho_{1}(\tau_{1},t)$ to Eq. (\ref{eq:f1differentialeq})
is represented by the connecting and the initial condition as

\begin{equation}
\rho_{1}(\tau_{1},t)=\begin{cases}
\Phi_{1}(t-\tau_{1})\exp\left(-\int_{0}^{\tau_{1}}\alpha(\tau)d\tau\right) & t>\tau_{1}\\
\rho_{1}(\tau_{1}-t,0)\exp\left(-\int_{\tau_{1}-t}^{\tau_{1}}\alpha(\tau)d\tau\right) & t<\tau_{1}
\end{cases}.
\end{equation}
Plugging the solution into the flux $\Phi_{0}(t)$, we obtain

\begin{align}
\Phi_{0}(t) & =\int_{0}^{t}\Phi_{1}(t-\tau_{1})\alpha(\tau_{1})\exp\left(-\int_{0}^{\tau_{1}}\alpha(\tau)d\tau\right)d\tau_{1}\nonumber \\
 & +\int_{t}^{\infty}\rho_{1}(\tau_{1}-t,0)\alpha(\tau_{1})\exp\left(-\int_{\tau_{1}-t}^{\tau_{1}}\alpha(\tau)d\tau\right)d\tau_{1},\label{eq:26}
\end{align}
where the first and the second terms in the right-hand side relate
to the connecting and the initial condition, respectively.

The integral-differential equations in the extended compartmental
SIS model \citep{Du_2020reliability} are obtained by representing
the recovery rate as the PDF of the infection duration (recovery time)

\begin{equation}
\psi_{R}(\tau_{I})=\alpha(\tau_{I})\exp\left(-\int_{0}^{\tau_{I}}\alpha(\tau)d\tau\right).
\end{equation}
We assume all the infected individuals get infected at the initial
time with the initial condition

\begin{equation}
\rho_{1}(\tau_{1},0)=\delta(\tau_{1})P_{1}(0).
\end{equation}
Then, the flux by Eq. (\ref{eq:26}) is rewritten as

\begin{equation}
\Phi_{0}(t)=\int_{0}^{t}\psi_{R}(\tau_{1})\Phi_{1}(t-\tau_{1})d\tau_{1}+\psi_{R}(t)P_{1}(0),
\end{equation}
which is the integral-differential equation in the extended compartmental
model \citep{Du_2020reliability}.

\subsection{The standard compartmental SIS model}

The standard compartmental SIS model is recovered by further assuming
the constant recovery rate $\alpha(\tau_{1})=\alpha$. The flux $\Phi_{0}(t)$
is simplified from Eq. (\ref{eq:Phi0}) as $\Phi_{0}(t)=\alpha P_{1}(t)$.
Together with Eq. (\ref{eq:Phi1constantbeta}), the ordinary differential
equations of the populations \citep{Brauer2019} follow as 
\begin{equation}
\dot{N}_{0}(t)=\alpha N_{1}(t)-\frac{\left\langle k\right\rangle \beta}{N_{T}}N_{0}(t)N_{1}(t)
\end{equation}
 and 
\begin{equation}
\dot{N}_{1}(t)=-\alpha N_{1}(t)+\frac{\left\langle k\right\rangle \beta}{N_{T}}N_{0}(t)N_{1}(t),
\end{equation}
with $N_{i}(t)=P_{i}(t)N_{T}$, $i=0,1$ as the susceptible and the
infected populations.

\section{The steady state in the mesoscopic model\label{sec:The-steady-state}}

In this section, we solve the steady state of the SIS model in the
mesoscopic model. In the steady state $\partial\rho_{k,i}(\tau_{i},t)/\partial t=0$,
the populations $n_{k,i}$ of the $k$-degree nodes in the state $i$
remains unchanged with the equal fluxes $\Phi_{k,0}=\Phi_{k,1}=\Phi_{k}$.
The equations of the steady-state coarse-grained PDFs are obtained
from Eqs. (\ref{eq:DDFof SISmoel})-(\ref{eq:Phik0}) as

\begin{align}
\frac{\partial\rho_{k,0}(\tau_{0})}{\partial\tau_{0}} & =-\Gamma_{k,0}(\tau_{0})\rho_{k,0}(\tau_{0})\label{eq:steadyfk0}\\
\Gamma_{k,0}(\tau_{0}) & =k\int_{0}^{\infty}\beta(\tau_{0},\tau_{1})\sum_{k^{\prime}=1}^{\infty}P(k^{\prime}|k)\rho_{k^{\prime},1}(\tau_{1})d\tau_{1}\label{eq:steadyGammak0}\\
\frac{\partial\rho_{k,1}(\tau_{1})}{\partial\tau_{1}} & =-\alpha(\tau_{1})\rho_{k,1}(\tau_{1})\label{eq:steadyfk1}\\
\Phi_{k,1} & =\int_{0}^{\infty}\Gamma_{k,0}(\tau_{0})\rho_{k,0}(\tau_{0})d\tau_{0}\\
\Phi_{k,0} & =\int_{0}^{\infty}\alpha(\tau_{1})\rho_{k,1}(\tau_{1})d\tau_{1},\label{eq:steadyPhik0}
\end{align}
with the connecting condition $\rho_{k,i}(0)=\Phi_{k,i},\;i=0,1$.

The steady-state solutions follow explicitly as

\begin{equation}
\rho_{k,0}(\tau_{0})=\Phi_{k}\exp[-\int_{0}^{\tau_{0}}\Gamma_{k,0}(\tau)d\tau],\label{eq:solutionfk0}
\end{equation}
and

\begin{equation}
\rho_{k,1}(\tau_{1})=\Phi_{k}\exp[-\int_{0}^{\tau_{1}}\alpha(\tau)d\tau].\label{eq:fk1}
\end{equation}
The steady-state probability follow as

\begin{equation}
P_{k,0}=\Phi_{k}\int_{0}^{\infty}\exp[-\int_{0}^{\tau_{0}}\Gamma_{k,0}(\tau)d\tau]d\tau_{0},\label{eq:nk0steady}
\end{equation}
and

\begin{equation}
P_{k,1}=\Phi_{k}\bar{\tau}_{1},\label{eq:nk1steady}
\end{equation}
where $\bar{\tau}_{1}$ is the average infection duration

\begin{equation}
\bar{\tau}_{1}=\int_{0}^{\infty}\exp[-\int_{0}^{\tau_{1}}\alpha(\tau)d\tau]d\tau_{1}.
\end{equation}
The conservation of the probability $P_{k,0}+P_{k,1}=1$ determines
the steady-state flux as

\begin{equation}
\Phi_{k}=\frac{1}{\int_{0}^{\infty}\{\exp[-\int_{0}^{\tau^{\prime}}\Gamma_{k,0}(\tau)d\tau]+\exp[-\int_{0}^{\tau^{\prime}}\alpha(\tau)d\tau]\}d\tau^{\prime}}.
\end{equation}

\subsection{Steady state in uncorrelated network}

In an uncorrelated network, the degree correlation is independent
of the degree $k$ as \citep{MarcBarthelemy2012}
\begin{equation}
P(k^{\prime}|k)=\frac{k^{\prime}P(k^{\prime})}{\left\langle k\right\rangle }.
\end{equation}
The transformation rate by Eq. (\ref{eq:steadyGammak0}) is simplified
as $\Gamma_{k,0}(\tau_{0})=k\Theta(\tau_{0})$, where the quantity
$\Theta(\tau_{0})$ is determined as

\begin{equation}
\Theta(\tau_{0})=\sum_{k^{\prime}=1}^{\infty}\int_{0}^{\infty}\beta(\tau_{0},\tau_{1})\frac{k^{\prime}P(k^{\prime})}{\left\langle k\right\rangle }\rho_{k^{\prime},1}(\tau_{1})d\tau_{1}.\label{eq:giveTheta}
\end{equation}
Therefore, the solution by Eq. (\ref{eq:solutionfk0}) is simplified
as

\begin{equation}
\rho_{k,0}(\tau_{0})=\Phi_{k}\exp[-k\int_{0}^{\tau_{0}}\Theta(\tau)d\tau].\label{eq:fk0steady}
\end{equation}
The steady-state flux, in turn, is rewritten as

\begin{equation}
\Phi_{k}=\frac{1}{\int_{0}^{\infty}\{\exp[-k\int_{0}^{\tau^{\prime}}\Theta(\tau)d\tau]+\exp[-\int_{0}^{\tau^{\prime}}\alpha(\tau)d\tau]\}d\tau^{\prime}}.\label{eq:Phiksteady}
\end{equation}
Equation (\ref{eq:giveTheta}) gives the self-consistent equation
for $\Theta(\tau_{0})$ as

\begin{align}
\Theta(\tau_{0}) & =\sum_{k^{\prime}=1}^{\infty}\frac{k^{\prime}P(k^{\prime})}{\left\langle k\right\rangle }\frac{\int_{0}^{\infty}\beta(\tau_{0},\tau_{1})\exp[-\int_{0}^{\tau_{1}}\alpha(\tau)d\tau]d\tau_{1}}{\int_{0}^{\infty}\{\exp[-k^{\prime}\int_{0}^{\tau^{\prime}}\Theta(\tau)d\tau]+\exp[-\int_{0}^{\tau^{\prime}}\alpha(\tau)d\tau]\}d\tau^{\prime}}.\label{eq:Thetasteady}
\end{align}

\subsection{Simple infection rate $\beta(\tau_{0},\tau_{1})=\beta(\tau_{1})$}

In the following, we consider the case that the infection rate $\beta(\tau_{0},\tau_{1})=\beta(\tau_{1})$
only depends on the infection duration $\tau_{1}$. The right-hand
side of Eq. (\ref{eq:Thetasteady}) does not rely on the susceptible
duration $\tau_{0}$, which results in a constant quantity $\Theta(\tau_{0})=\Theta$.
The integral on the right-hand side is worked out as $\int_{0}^{\infty}\{\exp[-k^{\prime}\int_{0}^{\tau^{\prime}}\Theta(\tau)d\tau]d\tau^{\prime}=1/(k^{\prime}\Theta)$.
Equation (\ref{eq:Thetasteady}) is simplified into Eq. (11) in the
main content.

The non-zero solution $\Theta$ exists for $\Upsilon>\Upsilon_{c}$,
where $\Upsilon_{c}=\left\langle k\right\rangle /\left\langle k^{2}\right\rangle $
is the epidemic threshold determined by the network structure. The
proof is given as follows.

We define a new function as
\begin{equation}
y(x)=1-\frac{\Upsilon}{\left\langle k\right\rangle }\sum_{k=1}^{\infty}\frac{k^{2}P(k)}{1+kx\bar{\tau}_{1}}.
\end{equation}
This function $y(x)$ is continuous and monotonously increasing for
$x>0$ with $\underset{x\rightarrow\infty}{\lim}y(x)>0$. The existence
of the positive solution to $y(x)=0$ requires $y(0)<0$, namely

\begin{equation}
1-\frac{\Upsilon}{\left\langle k\right\rangle }\sum_{k=1}^{\infty}k^{2}P(k)<0.
\end{equation}
The critical value gives the epidemic threshold $\Upsilon_{c}=\left\langle k\right\rangle /\left\langle k^{2}\right\rangle $.

For the large-size scale-free network with the degree distribution
$P(k)\propto k^{-\gamma},\:2<\gamma\leq3$, the divergence of the
second moment of the degree $\left\langle k^{2}\right\rangle =\sum_{k=1}^{\infty}k^{2}P(k)$
leads to zero epidemic threshold $\Upsilon_{c}=0$ of a large scale-free
network \citep{Pastor-Satorras2001}.

The fraction of the infected nodes is defined as

\begin{align}
r_{1}(t) & =\frac{\sum_{k=1}^{\infty}n_{k,1}(t)}{\sum_{k=1}^{\infty}n_{k}}\label{eq:fractionofinfectednodesr1}\\
 & =\sum_{k=1}^{\infty}P(k)P_{k,1}(t).
\end{align}
With the steady-state probability $P_{k,1}$ by Eq. (\ref{eq:nk1steady}),
the steady-state fraction of the infected nodes is

\begin{equation}
r_{1}=\sum_{k=1}^{\infty}\frac{k\Theta\bar{\tau}_{1}}{1+k\Theta\bar{\tau}_{1}}P(k),\label{eq:fractionsteadystate}
\end{equation}
which is positive with $\Theta>0$.

\section{Continuous-time Monte Carlo Simulation of the SIS Model\label{sec:Monte-Carlo-Simulation}}

This section shows the numerical simulation of the duration-dependent
SIS model on networks. In the previous studies \citep{Li2012}, the
simulation of the duration-dependent model is formulated by recording
all the possible events in the timeline, referred to as the tickets.
The states of the nodes are updated through the tickets. New tickets
are generated from infected nodes. In our algorithm, instead of recording
the tickets which may or may not occur, we only record the final time
when the node will leave the current state, which saves the memory
and gives the same results.

\subsection{Simulation algorithm \label{subsec:Simulation-algorithm-1}}

\begin{figure}
\includegraphics[width=11cm]{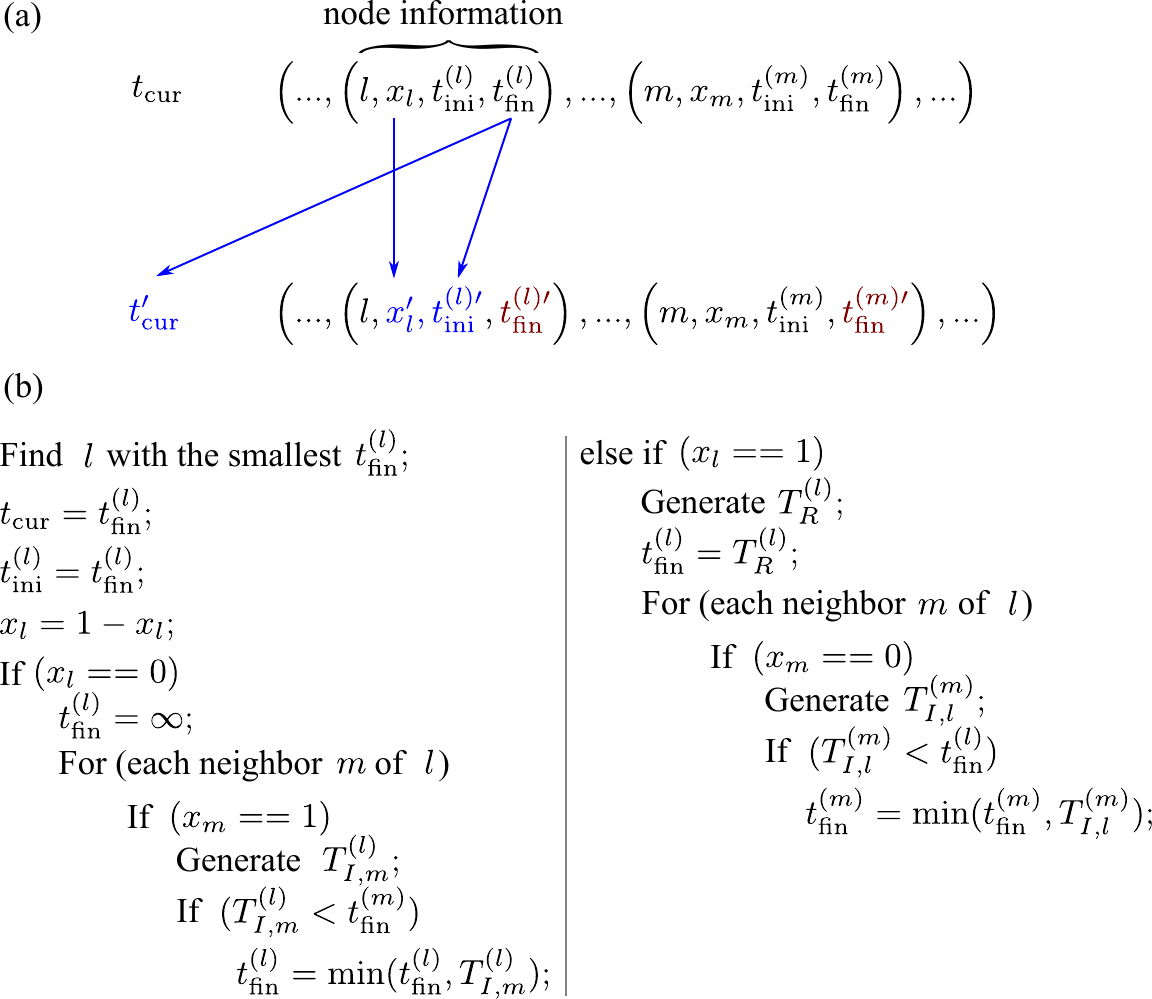}

\caption{One-step evolution in the simulation. (a) The updated data in the
one-step evolution. The parameters with prime represent the renewed
parameters. Finding the node $l$ with the smallest final time $t_{\mathrm{fin}}^{(l)}$,
the next event is executed by updating the current time as $t_{\mathrm{cur}}^{\prime}=t_{\mathrm{fin}}^{(l)}$
and the state and the initial time of the node $l$ as $x_{l}^{\prime}=1-x_{l}$
and $t_{\mathrm{ini}}^{(l)\prime}=t_{\mathrm{fin}}^{(l)}$. The new
final time $t_{\mathrm{fin}}^{(l)\prime}$ is then generated according
to the basic process. This event might also affect the final time
$t_{\mathrm{fin}}^{(m)\prime}$ of the neighbor $m$. (b) The pseudo
code of the one-step evolution.\label{fig:One-step-updating-of}}
\end{figure}

The current time $t_{\mathrm{cur}}$ represents the time of the current
step. For each node, we record the state of the node $x_{l}$, the
initial time $t_{\mathrm{ini}}^{(l)}$ when the node transformed to
the current state, and the final time $t_{\mathrm{fin}}^{(l)}$ when
the node will transform to the other state, as shown in Fig. \ref{fig:One-step-updating-of}
(a). The susceptible and the infected states are $x_{l}=0$ and $x_{l}=1$.
At the beginning, the current time $t_{\mathrm{cur}}$ is set as $0$.
The state of the network is prepared by assigning the state $x_{l}$
for each node. The initial time $t_{\mathrm{ini}}^{(l)}$ and the
final time $t_{\mathrm{fin}}^{(l)}$ for each node are set as $t_{\mathrm{ini}}^{(l)}\leq0$
and $t_{\mathrm{fin}}^{(l)}>0$, respectively.

With the prepared state, the evolution of the spread is realized step
by step. In each step, an event occurs with the state change of one
node. There are two kinds of events in the SIS model: the recovery
(infection) event with a node transforming from the state $1$ $(0)$
to the state $0$ $(1)$. Since the future events are recorded by
the final time of the nodes, the next event is obtained by finding
the node $l$ with the smallest final time $t_{\mathrm{fin}}^{(l)}$.
We give the explicit procedure of the updating for the recovery and
the infection event as follows. The pseudo code is shown in Fig. \ref{fig:One-step-updating-of}
(b).

For either a recovery or an infection event, the current time is updated
with the smallest final time as $t_{\mathrm{cur}}^{\prime}=t_{\mathrm{fin}}^{(l)}$,
which records the time of the current event and prepares for the next
event. The new state of the node $l$ is $x_{l}^{\prime}=1-x_{l}$
with the new initial time $t_{\mathrm{ini}}^{(l)\prime}=t_{\mathrm{fin}}^{(l)}$,
as shown in Fig. \ref{fig:One-step-updating-of} (a). In a recovery
event, the node $l$ recovers to the susceptible state $x_{l}^{\prime}=0$,
and may get infected again from an infected neighbor in the following
evolution. The new final time is first set as $t_{\mathrm{fin}}^{(l)\prime}=\infty$,
and is then updated according to the infection time generated from
the infected neighbors. For each infected neighbor $m$ of the node
$l$, an infection time $T_{I,m}^{(l)}$ is generated according to
the accumulated distribution as

\begin{equation}
\mathrm{Pr}(t_{\mathrm{cur}}^{\prime}<T_{I,m}^{(l)}<t_{I,m}^{(l)})=1-\exp[-\int_{t_{\mathrm{cur}}^{\prime}}^{t_{I,m}^{(l)}}\beta(t-t_{\mathrm{ini}}^{(m)})dt].
\end{equation}
The infection time $T_{I,m}^{(l)}$ is valid when it is smaller than
the final time $t_{\mathrm{fin}}^{(m)\prime}$ of the infected neighbor
$m$. If at least one valid infection time exists, the new final time
$t_{\mathrm{fin}}^{(l)\prime}$ of the node $l$ is updated as the
smallest valid infection time.

In an infection event, the node $l$ gets infected $x_{l}^{\prime}=1$.
The new final time $t_{\mathrm{fin}}^{(l)\prime}$ is generated as
the recovery time $T_{R}^{(l)}$ according to the accumulated distribution
as

\begin{equation}
\mathrm{Pr}(t_{\mathrm{cur}}^{\prime}<T_{R}^{(l)}<t_{R}^{(l)})=1-\exp[-\int_{t_{\mathrm{cur}}^{\prime}}^{t_{R}^{(l)}}\alpha(t-t_{\mathrm{cur}}^{\prime})dt].
\end{equation}
The new infected node $l$ may infect his neighbor in the future.
The final time of the susceptible neighbors of the node $l$ may change.
For each susceptible neighbor $m^{\prime}$, an infection time $T_{I,l}^{(m^{\prime})}$
is generated according to the accumulated distribution as

\begin{equation}
\mathrm{Pr}(t_{\mathrm{cur}}^{\prime}<T_{I,l}^{(m^{\prime})}<t_{I,l}^{(m^{\prime})})=1-\exp[-\int_{t_{\mathrm{cur}}^{\prime}}^{t_{I,l}^{(m^{\prime})}}\beta(t-t_{\mathrm{cur}}^{\prime})dt].
\end{equation}
If the infection time $T_{I,l}^{(m^{\prime})}$ is smaller than the
new final time $t_{\mathrm{fin}}^{(l)\prime}$ of the node $l$, the
final time $t_{\mathrm{fin}}^{(m^{\prime})\prime}$ of the susceptible
neighbor $m^{\prime}$ is updated as the earlier one between itself
and the infection time $T_{I,l}^{(m^{\prime})}$.

\subsection{Transition Rate of Weibull distribution}

In the simulation, we consider the recovery and the infected duration
satisfy the Weibull distribution. The cumulative distribution function
of Weibull distribution is

\begin{equation}
\mathrm{Pr}(0<T<t)=1-\exp[-(t/b)^{a}],
\end{equation}
which gives the transition rate

\begin{align}
\alpha_{\mathrm{W}}(t) & =\frac{\frac{d}{dt}\mathrm{Pr}(0<T<t)}{\mathrm{Pr}(T\geq t)}\nonumber \\
 & =\frac{a}{b}\left(\frac{t}{b}\right)^{a-1}.
\end{align}
The Weibull distribution returns to the exponential one with $a=1$.
In the following simulation of the SIS model, the recovery and the
infection rates are chosen as

\begin{equation}
\alpha(\tau_{1})=\frac{a_{\alpha}}{b_{\alpha}}\left(\frac{\tau_{1}}{b_{\alpha}}\right)^{a_{\alpha}-1}
\end{equation}
and

\begin{equation}
\beta(\tau_{1})=\frac{a_{\beta}}{b_{\beta}}\left(\frac{\tau_{1}}{b_{\beta}}\right)^{a_{\beta}-1}.
\end{equation}

\subsection{Generating the uncorrelated Scale-free network}

The uncorrelated scale-free network is generated by the configuration
model \citep{Catanzaro2005} for $N_{T}=2500$ nodes. The numbers
of the $k$-degree nodes are set as approximation integers

\begin{equation}
n_{k}=\frac{1/k^{3}}{\sum_{k^{\prime}=k_{\mathrm{min}}}^{k_{\mathrm{max}}}1/k^{\prime3}}N_{T},
\end{equation}
with $k$ ranging from the minimal degree $k_{\mathrm{min}}=10$ to
the maximal degree $k_{\mathrm{max}}=50$. The maximal degree is set
as $k_{\mathrm{max}}\leq\sqrt{N_{T}}=50$ to ensure an uncorrelated
network \citep{Catanzaro2005}. With the assigned degree for each
node, all nodes are randomly linked avoiding multiple and self-connection.
For an uncorrelated network, the degree correlation is determined
by the degree distribution as $P(k^{\prime}|k)=k^{\prime}P(k^{\prime})/\left\langle k\right\rangle $.

\subsection{Simulation results of single run}

\begin{figure}
\includegraphics[width=7cm]{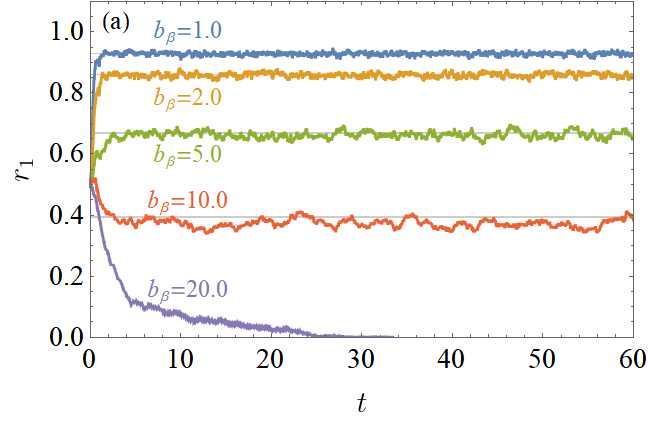}\includegraphics[width=7cm]{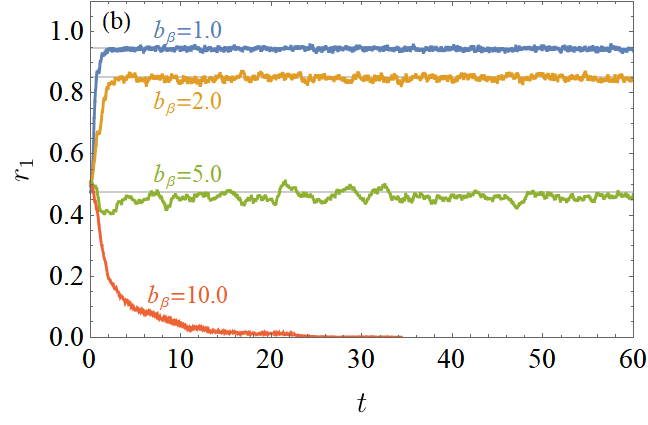}

\caption{The fraction $r_{1}(t)$ of the infected nodes in single run. The
recovery and the infection rates are chosen as $\alpha(\tau_{1})=1$
and $\beta(\tau_{1})=a_{\beta}/b_{\beta}(\tau_{1}/b_{\beta})^{a_{\beta}-1}$
with $a_{\beta}=1$ in (a) and $a_{\beta}=1.5$ in (b). The colored
curves present the simulation results, and the gray horizontal lines
present the steady-state fraction of the infected nodes by Eq. (\ref{eq:fractionsteadystate}).
\label{fig:The-fraction-of}}
\end{figure}

We apply the simulation algorithm to simulate the spreading dynamics
of the SIS model in the uncorrelated scale-free network. Figure \ref{fig:The-fraction-of}
presents the simulation results (colored curves) of the fraction $r_{1}(t)$
of the infected nodes in single runs. The recovery and the infection
rates are chosen as $\alpha(\tau_{1})=1$ and $\beta(\tau_{1})=a_{\beta}/b_{\beta}(\tau_{1}/b_{\beta})^{a_{\beta}-1}$
with $a_{\beta}=1$ in (a) and $a_{\beta}=1.5$ in (b). For the initial
state, each node is randomly prepared in the state $x_{l}=0$ or $1$.
The initial time $t_{\mathrm{ini}}^{(l)}$ for each node is set as
$0$, and the final time $t_{\mathrm{fin}}^{(l)}$ is randomly set
between $0$ and $1$. After enough time of evolution, the system
reaches the steady state with $r_{1}(t)$ approaching the steady-state
fraction of the infected nodes by Eq. (\ref{eq:fractionsteadystate})
(gray horizontal lines). Due to the finite-size effect of the network,
$r_{1}(t)$ has some fluctuations in the steady state. Large fluctuation
appears for larger $b_{\beta}$ with smaller $\Upsilon$. For the
increasing $b_{\beta}$, the steady-state fraction $r_{1}$ of the
infected nodes decreases, and finally approaches zero with the refined
spreading rate satisfied $\Upsilon\leq\Upsilon_{c}$. For $b_{\beta}=20$
in (a) and $10$ in (b), the refined spreading rates are $\Upsilon=0.050$
and $0.042$ respectively, smaller than the epidemic threshold $\Upsilon_{c}=0.051$,
and the system finally reaches the disease-free state.

\bibliographystyle{apsrev4-1}
\bibliography{covidbib}